%% file: main.tex
\documentclass{nature}

\usepackage{xr}
\usepackage[symbol,perpage]{footmisc}
\usepackage{fixfoot}

\usepackage{etoolbox}
\makeatletter
\patchcmd\@fixed@footnote
  {\protected@xdef\@thefnmark{\csname @#1@fftn@footnote\endcsname}}
  {\protected@xdef\@thefnmark{%
     \expandafter\@fnsymbol\csname @#1@fftn@footnote\endcsname}}
  {}{}
\makeatother
\DeclareFixedFootnote{\repnote}{ These authors contributed equally to this work}

\title{Twin-Boundary Structural Phase Transitions in Elemental Titanium}

\author{Mohammad S. Hooshmand$^{1}$\repnote\, Ruopeng Zhang$^{1,2}$\repnote, Yan Chong$^{1,2}$, Enze Chen$^{1}$, Timofey Frolov$^{4}$, David L. Olmsted$^{1}$,  Andrew M. Minor$^{1,2,3}$  \& Mark Asta$^{1,3}$\footnote[2]{Email: \href{mailto: mdasta@berkeley.edu}{mdasta@berkeley.edu}}
  }

\usepackage{xr-hyper}
\usepackage{lineno,hyperref,ulem}
\usepackage{bigints}
\usepackage{rotating, adjustbox} 
\usepackage{import}
\usepackage{chapterbib}
\usepackage{color,soul}
\usepackage[symbol]{footmisc}

\makeatletter
\newcommand*{\addFileDependency}[1]{
  \typeout{(#1)}
  \@addtofilelist{#1}
  \IfFileExists{#1}{}{\typeout{No file #1.}}
}
\makeatother

\usepackage[table]{xcolor}
\definecolor{lightgray}{gray}{0.9}
\usepackage{multirow}

\usepackage{fancyhdr}
\usepackage{amssymb}
 \usepackage{amsmath}
\usepackage{amsfonts}

\usepackage[font=footnotesize,labelfont=bf]{caption}

 \usepackage{epsfig,graphicx,amsmath,txfonts}
\usepackage{float,afterpage,wrapfig}
\usepackage{multirow,tabularx,xcolor,colortbl}
\usepackage{tikz}
\usetikzlibrary{arrows,shapes,positioning,shadows,trees}
\usepackage{sidecap}
\usepackage{placeins }   
\usepackage{amsmath}

\usepackage{gensymb}

\modulolinenumbers[5]

\newcommand*{\et}[0]{\textit{et~al. }}
\newcommand*{\fig}[1]{Fig.~\ref{fig:#1}}
\newcommand*{\Fig}[1]{Figure~\ref{fig:#1}}

\newcommand*{\figext}[1]{Extended Data Fig.~\ref{fig:#1}}
\newcommand*{\tabext}[1]{Extended Data Table~\ref{tab:#1}}

\newcommand*{\figsup}[1]{Supplementary Fig.~\ref{figsup:#1}}
\newcommand*{\tabsup}[1]{Supplementary Table~\ref{tabsup:#1}}
\newcommand*{\notesup}[1]{Supplementary Note~\ref{secsup:#1}}
\newcommand*{\figsups}[1]{Supplementary Figs.~\ref{figsup:#1}}

\newcommand*{\tb}[0]{$\{11\bar{2}4\}$ }
\newcommand*{\tbBCO}[0]{$\text{BCO}~ \{11\bar24\}$ }
\newcommand*{\tbshuff}[0]{$\text{shuffle}~ \{11\bar24\}$ }
\newcommand*{\tbBCOO}[0]{$\text{BCO}~ \{11\bar24\}$}
\newcommand*{\tbshufff}[0]{$\text{Shuffle}~ \{11\bar24\}$}

\newcommand*{\eqn}[1]{Eq.~(\ref{eqn:#1})}

\newcommand*{\blue}[1]{\textcolor{black}{#1} }

\usepackage{float}
\floatstyle{plaintop}
\restylefloat{table}

\usepackage{graphicx}
\makeatletter
\let\saved@includegraphics\includegraphics
\AtBeginDocument{\let\includegraphics\saved@includegraphics}
\renewenvironment*{figure}{\@float{figure}}{\end@float}
\makeatother

\usepackage{chapterbib}

\newcommand{\beginsupplement}{%
        \setcounter{table}{0}
        \renewcommand{\tablename}{Supplementary Tab.}%

        \setcounter{figure}{0}
        \renewcommand{\figurename}{Supplementary Fig.}%
     }

\begin{document}

\input{Twin_Boundary_Structural_Phase_Transitions_in_Elemental_Titanium_rev}

\input{Twin_Boundary_Structural_Phase_Transitions_in_Elemental_Titanium_sup_rev}

\end{document}

%% file: Twin_Boundary_Structural_Phase_Transitions_in_Elemental_Titanium_rev.tex
\maketitle

\begin{affiliations}
 \item Department of Materials Science and Engineering, University of California, Berkeley, CA 94720, USA.
  \item National Center for Electron Microscopy, Molecular Foundry, Lawrence Berkeley National Laboratory, Berkeley, CA 94720, USA.
  \item Materials Sciences Division, Lawrence Berkeley National Laboratory, Berkeley, CA 94720, USA.
  \item Lawrence Livermore National Laboratory, Livermore, CA 94550, USA.

\end{affiliations}

\begin{abstract}

Twinning in crystalline materials plays an important role in many transformation and deformation processes, where underlying mechanisms can be strongly influenced by the structural, energetic and kinetic properties of associated twin boundaries (TBs).  While these properties are well characterized in common cases, the possibility that TBs can display multiple complexions with distinct properties, and phase transitions between them, has not been widely explored, even though such phenomena are established in a few more general grain boundaries.  We report experimental findings that \tb TBs in titanium display a thick interfacial region with crystalline structure distinct from the bulk.  First-principles calculations establish that this complexion is linked to a metastable polymorph of titanium, and exhibits behavior consistent with a solid-state wetting transition with compressive strain, and a first-order structural transition under tension.  The findings document rich TB complexion behavior in an elemental metal, with important implications for mechanical behavior and phase-transformation pathways.

\end{abstract}

Twinning is a commonly observed phenomenon associated with mechanical deformation, crystallization and phase transformations \cite{christian1995, sutton1995, kaplan2013, mishin2010}. Nucleation and growth of twins are strongly affected by the excess free energies and kinetic mobility of associated twin boundaries (TBs), i.e., the interfaces between twin variants. These interfacial properties are themselves strongly influenced by the details of TB interfacial atomic structure \cite{frolov2015, frolov2012, hart1972,  cahn1982, cantwell2020 }, and structure-property relationships for these interfaces have thus been the focus of extensive previous experimental and computational investigations \cite{mishin2010, meiners2020, frolovStructuralPhaseTransformations2013,frolov2013 }. Despite the detailed insights derived from previous works, the possibility of complexion transitions in TBs has not been widely explored. The term complexion has been introduced \cite{tang2006a, kaplan2013, dillon2007} to refer to the structure or ``phase” of an interface \cite{hart1968, hart1972, cahn1982}, associated with which are a unique set of interfacial excess properties and free energies \cite{frolov2012}. In grain boundaries, first-order transitions between competing complexions occur through changes in temperature and chemical potential (composition) (for recent reviews, see Refs. \cite{luo2016,cantwell2014,cantwell2020, kaplan2013 }).  Such transitions can strongly influence the interfacial thermodynamic properties \cite{gibbs1948, frolov2015, hart1972, cantwell2020}, and cause discontinuous changes in strength and mobility (e.g., \cite{frolov2014, cantwell2014, cantwell2020}) and other properties that influence microstructure, and macroscopic properties of polycrystalline materials \cite{dillon2007}.  \blue{While complexion transitions have been established in high-angle grain boundaries in several alloy systems \cite{duscher2004, frolov2015a, frolov2016, dillon2007}, fewer cases have been reported for elemental metals \cite{frolovStructuralPhaseTransformations2013, frolov2018, meiners2020, olmsted2011}. In particular, despite previous investigations into self-interstitial induced cluster formation in TBs of face-centered-cubic metals \cite{mendelev2013}, little is known about the nature of complexions and associated phase transitions in these interfaces more generally.} An understanding of the conditions under which complexion transitions may exist in TBs is of interest from the standpoint of materials design, given that these transitions can be controlled through solute additions and thermomechanical treatments to alter twinning processes that influence microstructure and properties.

In this work, we explore the interfacial structure of \tb TBs in hexagonal-close-packed (HCP) titanium ($\alpha$-Ti). These twins were reported to form in $\alpha$-Ti-O alloys deformed at cryogenic temperature, and were observed to display relatively thick nanometer-scale interfacial regions in electron-microscopy images \cite{chong2020}. Here we establish that this TB features a periodic atomic arrangement distinct from that of the bulk crystal structure.  Employing atomistic simulations coupled with evolutionary structural searches and first-principles calculations, we establish that this interfacial structure is an intrinsic property of elemental titanium, and can be linked to a strained metastable polymorph of this material.  \blue{The resulting complexion features a lower TB energy but higher excess volumes and interfacial stresses than a competing \tb TB structure found to be lower in energy in other HCP metals \cite{wangAtomicStructuresSymmetric2012}}. As a consequence, the TB properties are sensitive to strain state, and we demonstrate that applied strains can drive solid-state wetting and first-order complexion transitions. These findings have \blue{potentially important consequences for phase transformation pathways in general}, and the strong strain dependencies suggest interesting consequences for mechanical properties in $\alpha$-Ti.  More generally, the results provide insights into conditions where similar types of complexions may be realized in other systems, raising the possibility that their existence and effects may be discovered more broadly in a wider range of structural materials.

\subsection{Twin boundary structure:}
\label{sec:TB_structures}
 
\begin{figure}[h!]
 \begin{center}
   \includegraphics[width=\textwidth]{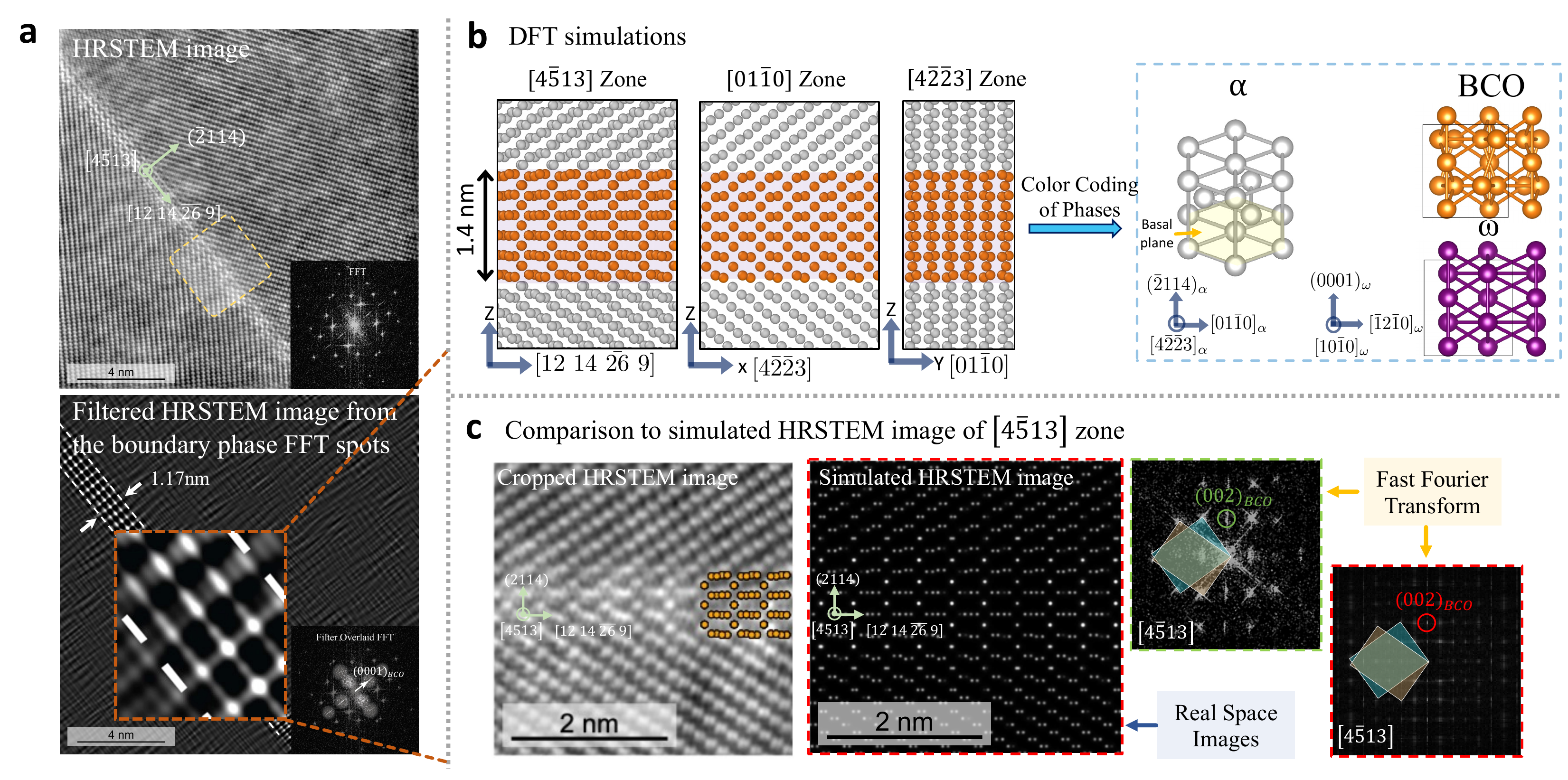}
\caption{Atomic structure of the \tb twin boundary in titanium. \textbf{a,} Experimentally measured HRSTEM image and filtered HRSTEM image from the boundary phase FFT spots for a \tb TB in a Ti-$0.3$ wt.\%O sample are shown on top and bottom, respectively . Insets show the FFT (overlaid with the applied filter in the lower inset). \textbf{b,} DFT calculated relaxed atomic structure of the \tb TB in pure Ti represented along different crystallographic zone axes with an extended region of the BCO structure at the interface (color codes are based on CNA parameter \cite{honeycutt1987} represented in the images next to these figures, \blue{in correspondence with the $[4\bar2\bar23]$ zone TB structure}, which also illustrate the close relation between the BCO structure and that of the well-known $\omega$ phase). \textbf{c,} Magnified HRSTEM and simulated HAADF HRSTEM image from DFT atomic positions from $[4\bar513]$ zone and FFT of each indicating the match with the BCO peaks along $[002]$ . The contrast in the thick boundary region in \textbf{a} is highlighted to demonstrate the matching of the measured and simulated images based on the DFT structures. }
  \label{fig:TEM}
 \end{center}
 \end{figure}

\Fig{TEM}a shows a high-resolution scanning transmission electron \blue{microscopy} (HRSTEM) image of the atomic structure of a \tb TB in a Ti-$0.3$ wt.\%O alloy that has been deformed at cryogenic temperatures \cite{chong2020}.  The image has been obtained through high-angle annular dark-field scanning transmission electron microscopy (HAADF-STEM), as described in the Methods section.  The structure features a relatively wide interfacial thickness compared to other common twin boundaries (cf., \figext{all_boundaries}). \blue{We observe that this interfacial width shows small fluctuations along the length of the TB, as will be discussed further below (and \notesup{more_exp})}. A Fast Fourier transform (FFT) of the HRSTEM image in the vicinity of the TB is plotted in the upper panel on the right side of \fig{TEM}c and shows peaks (cf., the peak circled in green) corresponding to the expanded interfacial region, which are located at positions distinct from those of the HCP twin variants (for higher resolution diffraction patterns, see \figext{FFT}). This observation suggests that there is well-ordered periodic structure formed at the TB which features an atomic arrangement distinct from the HCP phase of the bulk alloy.

To gain further insight into the nature of the interfacial structure we have undertaken atomistic and first-principles density-functional-theory (DFT) calculations to identify the lowest energy structure of the TB \blue{in elemental Ti (the role of oxygen will be discussed below).} The lowest-energy structure, with positions obtained from the DFT calculations, is shown in \fig{TEM}b in different projections, including the $[4\bar513]$ zone used in the HRSTEM image.  The atoms are color coded using a common-neighbor analysis (CNA) \cite{honeycutt1987}, with gray corresponding to the parent $\alpha$-Ti HCP structure, and orange corresponding to a body-centered orthorhombic (BCO) crystalline structure that is a strained version of a known metastable polymorph of elemental Ti \cite{zarkevich2016}. The BCO structure is related to the well-known $\omega$ polymorph of Ti, as shown in the panel on the right side of \fig{TEM}b.

A comparison is given of the experimental and computed structures in \fig{TEM}c.  The second panel from the left shows a simulation of an HRSTEM image formed from the DFT structure.  It is noted that both the experimental and simulated image feature similar structural motifs including alternating planes containing either a periodic array of ``bright” spots or closely-spaced atomic columns parallel to the interface, which are structural features not observed in the regions away from the TB \blue{(see \figsups{4DSTEM}-\ref{figsup:HRSTEM_width})}.  Further, the simulated FFT of the interfacial region derived from the calculations, shown in the lower panel on the right, contains extra spots at non-HCP positions consistent with the FFT of the experimental image.  These spots can be identified as originating from the $(001)$ planes of the strained BCO structure, with an interplanar spacing of $\sim 3.04 \, \AA$. We do note that the experimental TB structure is observed to be slightly narrower than calculated by DFT, a result we rationalize as being due to differences in the relative (free) energies between the HCP and BCO structures for the zero-temperature calculations for pure Ti and room-temperature experiments in $\alpha$-Ti-O alloys, as discussed in \blue{\notesup{disjoining}}. Overall, the results summarized in \fig{TEM}c demonstrate striking similarity between the measured and calculated structures, providing experimental confirmation of the calculated structure, the origins of which we analyze next.

\subsection{ Twin boundary energy and excess properties:}
\label{sec:excess}

\begin{figure}[h!]
 \begin{center}
   \includegraphics[width=\textwidth]{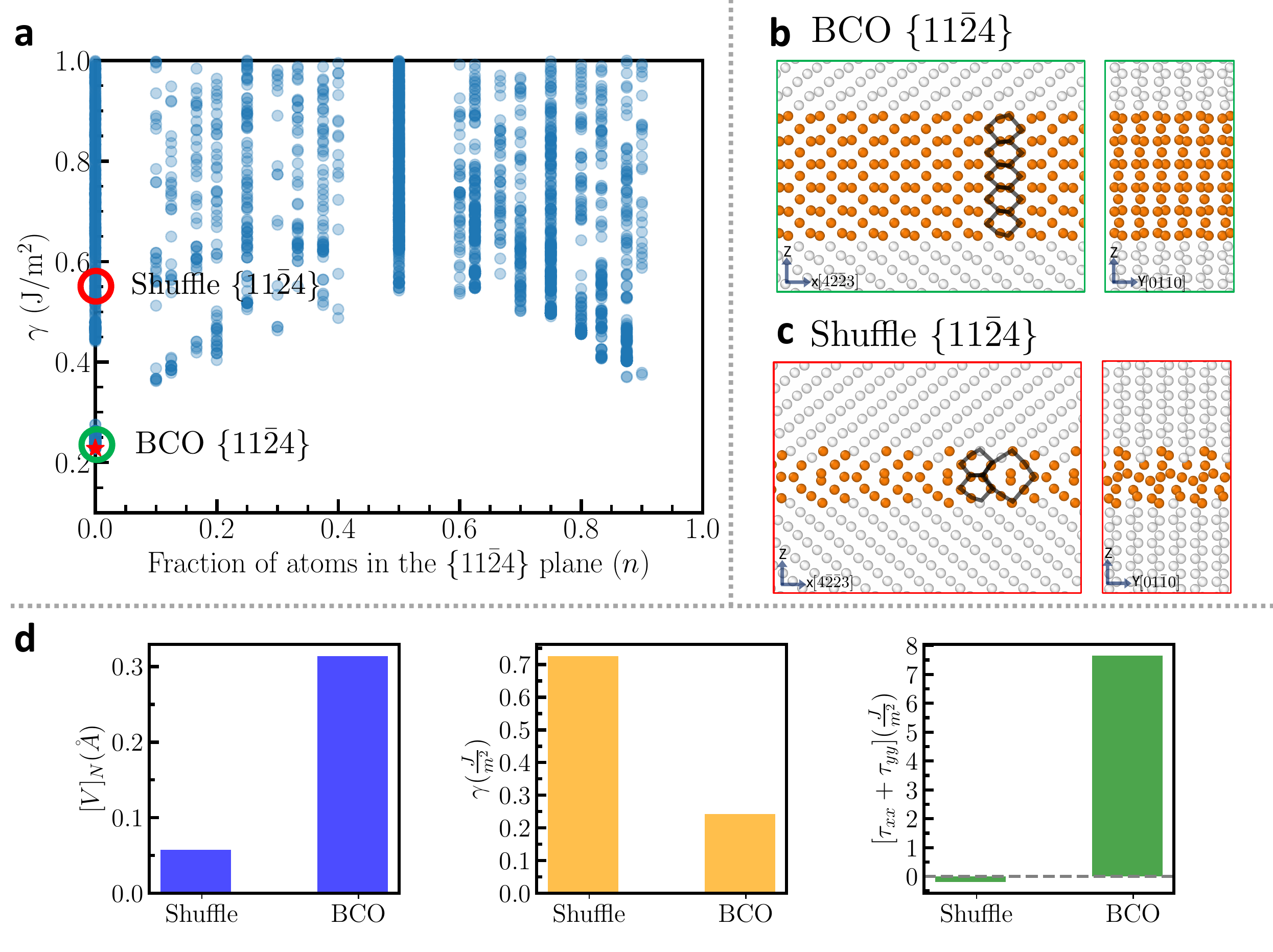}
\caption{ \textbf{a,} Grand-canonical structure search results for the \tb twin boundary in Ti. Twin boundary (TB) energies of stable and metastable structures are plotted versus the number of excess interfacial atoms ($n$), computed as a fraction of the number of atoms in the \tb bulk atomic plane as defined by Frolov \et \cite{frolovStructuralPhaseTransformations2013}. The red star at $n=0$ indicates that the ground state structure (labeled as \tbBCO) does not require the addition or removal of Ti atoms. \blue{Another competing state (labeled as \tbshuff) with a higher excess energy than the ground state structure is indicated by the red circle in the excess energy plot.} \textbf{b,} \textbf{c,} show the atomic structure of these two competing states after DFT relaxation. BCO structural units are illustrated with black lines, and appear in both TB structures, but with the higher energy TB showing an additional structural unit. Color codes are based on CNA parameter \cite{honeycutt1987} with gray representing HCP and orange showing defective regions. \textbf{d,} DFT-calculated values for the excess volume ($[V]_\text{N}$), excess energy ($\gamma$), and trace interfacial stress ($\tau_{xx}+\tau_{yy}$) are plotted for the \tbBCO and \tbshuff TBs, highlighting the significant differences in these interfacial excess quantities between the two TB complexions.} 
  \label{fig:unique}
 \end{center}
 \end{figure}

Further insight into the structure and energetic factors underlying the \tb TB structure for $\alpha$-Ti can be derived from the computational results summarized in \fig{unique}. The left panel shows the results of a search over \tb TB structures, performed using an evolutionary search algorithm \cite{zhu2018, oganov2006, lyakhov2013} based on the classical interatomic potential model of Hennig \et \cite{hennig2008} that has been fit to DFT energies and forces.  The results plot the interfacial energy of all stable and metastable structures identified by the evolutionary search as a function \blue{of} the fraction of atoms added/removed from the bulk \tb planes joined to form the TB. The green circle and red star identify the lowest-energy BCO TB structure, which will hereafter be referred to as ``\tbBCOO", which is again illustrated in \fig{unique}b (after further relaxation by DFT).  The results demonstrate that no addition or removal of atoms is required in the interfacial region to form the \tbBCO structure.  We also note in \fig{unique}a, that there are a few other points with energies very close to the \tbBCO  structure; these involve the same basic BCO interfacial atomic arrangement, but with slight distortions or relative displacements of the adjacent HCP crystals that raise the energy.  The \tbBCO  and these associated structures form a cluster of energy points in \fig{unique}a that are distinctly lower than all other identified TB structures, across the different values of the interfacial excess number of atoms. Also highlighted by the red circle in \fig{unique}a, is a higher-energy TB structure that we label ``\tbshufff", the DFT-minimized structure of which is shown in \fig{unique}c. \blue{Although this structure is considerably higher in energy for $\alpha$-Ti, we highlight it because it corresponds to a metastable structure with the same number of atoms and a structural motif distinct from the \tbBCO, and we find that it becomes lower in energy under tensile strain as discussed below. The crystallographic characteristics of different atomic structures of the \tb TB are detailed in \notesup{crystal}. }

Plotted in \fig{unique}d are the excess volume per area ($[V]_\text{N}$), excess interfacial energy ($\gamma$), and the trace of the interfacial stress ($\tau_{ij}= \frac{\partial \gamma}{\partial \varepsilon_{ij}}+\delta_{ij}\gamma$, where $\varepsilon_{ij}$ is the strain measured relative to the undeformed HCP structure, and $\delta_{ij}$ is the Kronecker delta). The values are tabulated in \tabext{excess} and plotted in \figext{energies} for \tb and other common TBs in Ti. Despite the lower value of $\gamma$ for the \tbBCO TB, it is seen to feature considerably higher excess volume and stress than the \tbshuff complexion.  The magnitude of ($\tau_{xx}+\tau_{yy}$) in particular is highly noteworthy for the \tbBCO TB since it is more than an order of magnitude larger than the value of the interfacial energy itself, a result we rationalize below through a model for the BCO complexion as one that involves wetting of the interface by the BCO polymorph of Ti that is under tensile stress.  The large values of the differences in excess volume and interfacial stress raise interesting possibilities for complexion transitions induced by strain, as we explore next.

\subsection{ Strain-induced complexion transitions:}
\label{sec:strain}

We explore this possibility using DFT calculations to compute the interfacial energy of the \tbBCO TB and \tbshuff complexions as a function of biaxial strain $\varepsilon_{\text{xx}}=\varepsilon_{\text{yy}}=\varepsilon$, where $x$ and $y$ refer to directions perpendicular and parallel to the tilt axis, respectively. In these calculations $\varepsilon$ is varied from $-2.5 \, \%$ (compression) to $7 \, \%$ (tension) and $\gamma$ as a function of $\varepsilon$ is computed from \eqn{GBE}.

Results of $\gamma$ versus $\varepsilon$ are plotted for the \tbBCO and \tbshuff TB complexions in \fig{transitions}a. \blue{These calculations are carried out using both free-surfaces (FS) and periodic boundary conditions (PBC) (see Methods)} The corresponding relaxed structures of the \tbBCO TB at representative strains are shown in \fig{transitions}b, highlighting changes in interfacial width induced by varying $\varepsilon$. We begin by considering tensile values of the strain and note that as $\varepsilon$ increases the value of $\gamma$ for the \tbBCO twin rises sharply, consistent with the large and tensile value of the interface stress.  By contrast, $\gamma$ for the \tbshuff TB shows a weak decrease with increasing $\varepsilon$ near zero strain, consistent with its small and negative interfacial stress, and decreases more rapidly at larger positive values of $\varepsilon$.  The two curves for $\gamma$ versus $\varepsilon$ cross at a biaxial strain of $\varepsilon=6.4 \, \%$, indicating a strain-induced interfacial phase transition from the \tbBCO to the \tbshuff complexion at this value of tensile strain.  \blue{We note that the transition is first order, such that there is not a continuous transformation between the two complexions, i.e., the BCO complexion does not spontaneously transform to the shuffle complexion for strains above $6.4\, \%$, and vice versa below $6.4\,\%$ (i.e., these these phases exist as metastable states above or below the transformation point). } This first-order transition can be triggered not only by imposing biaxial strain, but also through hydrostatic or normal stress. Specifically, from the Gibbs adsorption equation for interfaces in crystalline materials \cite{frolov2012, cantwell2020}, the negative of the excess volume gives the rate of change of $\gamma$ with applied stress $\sigma_{\text{zz}}$ normal to the interface: $\partial \gamma / \partial \sigma_{\text{zz}} = -[V]_N$.  The lower value of $[V_N]$ for the \tbshuff TB implies that a first-order transition from \tbBCO to this complexion can also be induced by applied compressive stress.  The combination of compressive normal stress and tensile strain parallel to the interface would then act synergistically to induce this transition at lower stresses/strains.

Compared to the behavior under tension, the behavior under compressive biaxial strain is qualitatively different.  As shown in \fig{transitions}a, application of compressive values of $\varepsilon$ leads initially to a decrease in $\gamma$ for the \tbBCO TB. However, as compressive strain is increased, the width of the interface increases and at an applied strain of approximately $\varepsilon=-1.4 \, \%$ we find that the entire supercell transforms to the BCO phase. This behavior is consistent with a solid-state wetting transition of the interface by the BCO polymorph structure.  Specifically, as discussed in the \blue{\notesup{disjoining}}, $\gamma$ for the \tbBCO TB can be understood as reflecting a competition between (i) the energy per area to transform a region of HCP structure to BCO (which is strained relative to bulk), \blue{which is proportional to the energy difference $\Delta E$ between the bulk (strained) BCO and HCP structures}, and (ii) the change in interface energy when a narrow TB structure is replaced by the two BCO/$\alpha$ interfaces separated by distance $w$.  The first term depends linearly on $w$ and the second generally non-linearly, as described by a \blue{so-called ``disjoining potential" (DP) \cite{lipowsky1987, fisher1985}. In this picture, for a fixed DP, the equilibrium value of $w$ will increase (decrease) with decreasing (increasing) values of $\Delta E$.}  As discussed in \blue{\notesup{disjoining}}, if the \blue{DP} features a weak attractive minimum, the equilibrium value of $w$ will increase with increasing compressive strain, which lowers \blue{$\Delta E$}, until a critical strain where the width diverges, consistent with the calculated results \blue{(see \figsup{wet})}.

\blue{Further support of the DP theoretical framework for understanding BCO complexion formation and wetting is provided by DFT calculations employing an alternate exchange-correlation functional (GGA+U) that leads to increased values of $\Delta E$ relative to the GGA functional used for the results in \fig{TEM}.  The same BCO complexion is predicted by GGA+U, but with a width that is smaller relative to GGA.  This result is consistent with predictions of the DP theory, due to the higher value of $\Delta E$ for GGA+U (see \notesup{exchange}).}

\blue{Further support for the DP picture is provided by the experimental observations that $w$ shows small variations along the TB width.  This would be consistent with the expected variations along the TB area in both the oxygen concentration, and the state of stress (strain) in the mechanically deformed sample.  Both types of variations are expected to change $\Delta E$, which in turn would lead to changes in $w$.  }

\blue{Finally, the DP formalism also provides a framework for understanding why the DFT results for pure Ti in \fig{TEM} predict an interfacial width larger than in the experimental sample with oxygen.  DFT calculations presented in \notesup{oxygen} show that oxygen displays no energetic preference to segregate to the \tb TB, and that it raises the energy difference between BCO and HCP structures.  Once again this effect should lead to smaller values of $w$ in samples with oxygen than in pure Ti, consistent with the results in \fig{TEM}.  The DFT calculations also suggest that the fact that the \tb TB was only observed to form in samples with high oxygen content (and low deformation temperatures) is not due to the oxygen-segregation to the TB, but rather due to the oxygen's role in hardening the material (see  \notesup{oxygen}). }

\subsection{\blue{Structural relationships between the polymorphs in $\alpha$-Ti:}}
\label{sec:omega}

\begin{figure}[h!]
 \begin{center}
   \includegraphics[width=\textwidth]{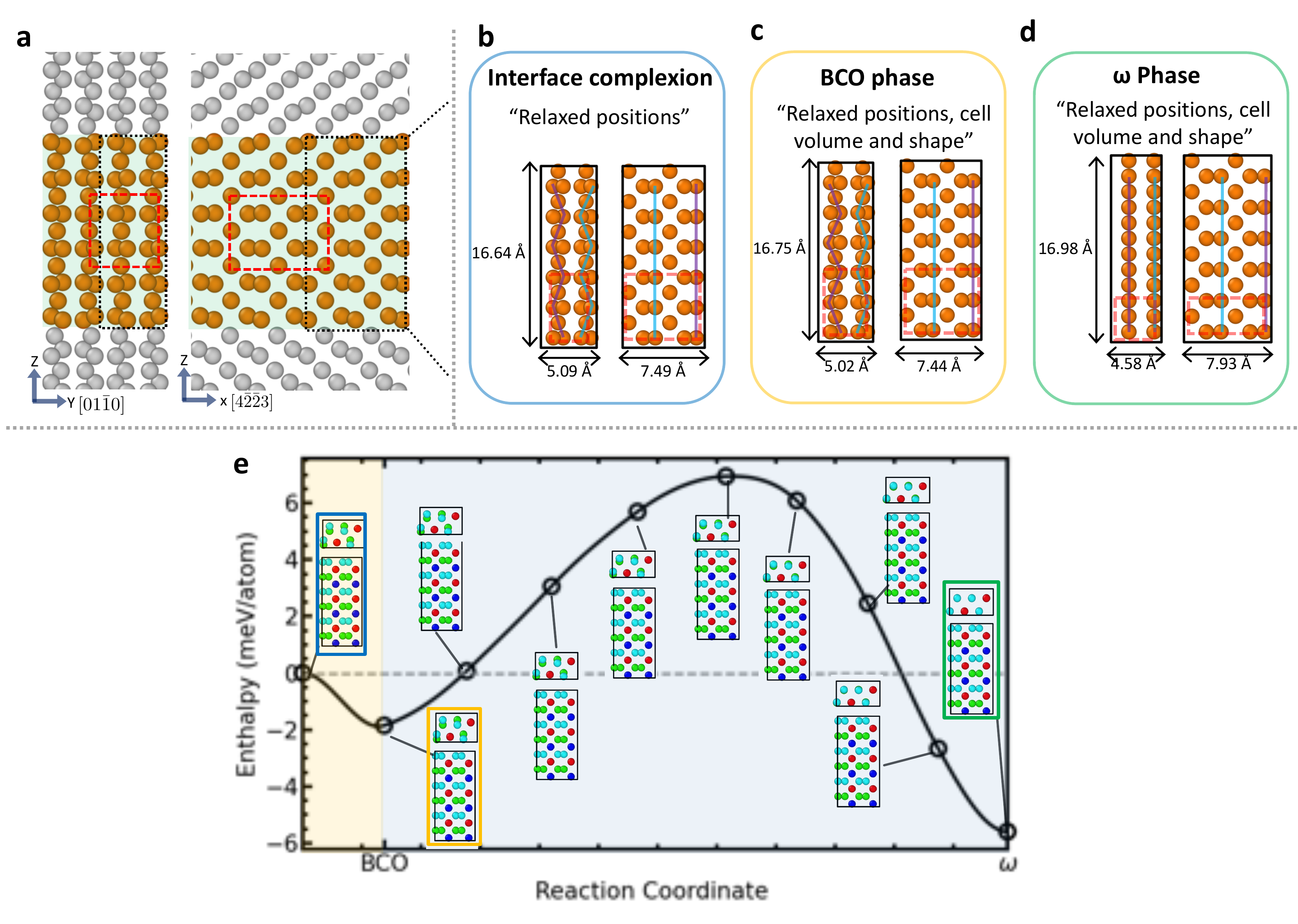}
\caption{Strained BCO to unstrained $\omega$ phase transformation pathway. \textbf{a,} Relaxed \tb twin boundary (periodic images are doubled in $x$ and $y$ directions as a guide for the eye). \textbf{b,} shows the as-taken extended interface region in which the atomic positions are relaxed by keeping the in-plane dimension fixed at the value corresponding to the interface and relaxing in the direction normal (strained BCO). \textbf{c,} shows the fully relaxed (atomic positions, cell volume and shape) cell which corresponds to the stress-free bulk BCO structure. \textbf{d,} shows the fully relaxed $\omega$ phase. \textbf{e,} Calculated minimum energy pathway (MEP) plotting the energy profile versus reaction coordinate of strained BCO relaxed to the strain-free BCO and $\text{BCO} \rightarrow \omega$ phase transition (insets show the $z$ (top) and $y$ (below) projection of atoms). Red and black dashed lines in \textbf{a} show the BCO conventional cell ($12$ atoms) and interfacial supercell ($36$ atoms) with a triple periodicity along $z$, respectively. Red dashed lines in \textbf{d} also show the $6$ atom conventional cell of the $\omega$ phase. Gray and orange atoms in \textbf{a-d} are HCP and interfacial atoms as determined by common neighbor analysis (CNA) \cite{honeycutt1987}, respectively. Atoms in \textbf{e} are colored based on their $z$ coordinates.}
  \label{fig:NEB}
 \end{center}
 \end{figure}

\blue{We consider next the relation between the interfacial BCO complexion identified at the \tb TB and other relevant polymorphs of bulk Ti.} As illustrated in \fig{NEB}a and b, we identify from the relaxed TB structure the smallest conventional orthorhombic unit cell as corresponding to the red box in \fig{NEB}a.  We form a bulk crystal structure based on this $12$-atom unit cell and perform a DFT geometrical relaxation keeping the unit cell dimensions parallel to the interface fixed at the dimensions found in the TB, while allowing the dimension normal to the interface to relax. The resulting structure is shown in \fig{NEB}b, which we will refer to in what follows as the ``interfacial BCO phase” and its energy provides a reference for the analysis that follows.  We then allow a full geometric relaxation in DFT, yielding the strain-free BCO phase shown in \fig{NEB}c. Comparing the final relaxed BCO structure to the interfacial BCO phase, we find that the latter is under a state of tensile strain, corresponding to $\varepsilon_{\text{xx}}=0.007$ and $\varepsilon_{\text{yy}}=0.014$ relative to the strain-free structure.  As indicated above, the fact that the interfacial BCO phase is under a state of tensile strain provides an explanation for the high values of the tensile interfacial stress computed for the \tbBCO TB. 

For comparison, we also show in \fig{NEB}d the structure of the $\omega$ phase, illustrating the correspondence with the BCO phase.  The orientation relationship for the $\omega$ phase relative to $\alpha$ in this representation is \blue{$(0001)_{\omega} \parallel (\bar{2}114)_{\alpha}$, $ [10\bar{1}0]_{\omega} \parallel [4\bar{2}\bar{2}3]_{\alpha}$}.  Comparing \fig{NEB}(c) and (d) highlights that the two structures are related by homogeneous strains and atomic shuffles.  Concerning the latter, we note that the periodicity of the atomic structure in the BCO phase repeats once every two $[0001]_{\omega}$ planes.

\blue{We next examine the relationship between these different phases by computing the minimum energy pathway (MEP) between the interfacial BCO and relaxed $\omega$ structures.} \Fig{NEB}e plots the energy versus reaction coordinate along the MEP. The (strained) interfacial BCO phase (corresponding to a zero value of the reaction coordinate) is connected by a barrierless path to the relaxed BCO phase (first minimum along the reaction path). \blue{This result is important because it establishes that the interfacial BCO phase is simply a strained version of the bulk BCO phase, and the lack of an energy barrier between them implies they are the same phase.  By contrast, the} relaxed BCO phase is separated by a $9.1 \, \text{meV}/\text{atom}$ barrier from the relaxed $\omega$ phase, across a MEP that accomplishes homogeneous strains of $\varepsilon_{\text{xx}}=0.06$, $\varepsilon_{\text{yy}}=-0.09$, and $\varepsilon_{\text{zz}}=0.01$, as well as a series of atomic shuffles, to transform from the relaxed BCO to the $\omega$ phase. \blue{This result shows that BCO and $\omega$ are indeed separate phases, separated by an energy barrier along the MEP.  Further, as shown by the DFT calculations in \figsup{DFV}, $\omega$ has a considerably higher energy than BCO under the biaxial strain state imposed in the \tb TB, such that there is no bulk driving force for $\omega$ to form at this interface.}

\section*{Outlook}
\label{sec:outlook}

The present work has identified an interfacial complexion in \tb TBs of $\alpha$-Ti that is characterized by a strong strain dependence of the interfacial free energy, and associated first-order interfacial phase transitions and solid-state wetting behavior under tensile and compressive biaxial loading, respectively. This behavior can be interpreted as arising from two key features of the interface complexion: (i) it displays a periodic crystalline structure that is linked by a barrierless energy path to a metastable polymorph of Ti (i.e., the BCO phase), and (ii) this polymorph has available to it a low-energy orientation relationship with $\alpha$-Ti, for the crystallography characteristic of the TB, leading to a low heterophase ($\alpha$/BCO) interface energy, at the expense of imposing a misfit strain on the polymorph (BCO) phase. We note that feature (i) is not uncommon and has been evidenced in $\beta-$Ti and other material systems \cite{Zhang2017, medlin2001}. However, what has not been reported, to the best of our knowledge, is the role that the strain required to accommodate such metastable polymorphs within the interface (i.e., ingredient (ii) above) plays in driving anomalous values of the excess interfacial quantities and associated complexion transitions. The presence of competing metastable polymorphs that are close in energy to the stable phase is common in many material systems, and the five-dimensional crystallographic space of interfaces provides numerous possibilities for accommodating these metastable structures with moderate misfit strains and low-energy orientation relationships. Therefore, formation of the class of polymorph complexions reported here, as well as the types of related interfacial phenomena identified in this work, could be expected to be encountered for other materials systems far more generally.

The findings also provide a framework for controlling features of interfaces that impact materials properties. For example, it has been observed experimentally that the formation of \tb deformation twins in Ti-O alloys correlates with fracture processes that underly a sharp decrease in ductility \cite{chong2020}. \blue{The correlation between observed cracks and the location of the \tb twins reported in Ref. \cite{chong2020} suggest that the large excess volumes and stresses reported here for this TB could enhance fracture and therefore contribute to embrittling the sample.}  Thus, destabilization of the BCO complexion of this TB in favor of an alternate interfacial phase that can realize the benefits of twinning (plastic deformation and work hardening), without the detrimental effects associated with large excess stresses, could be beneficial in enhancing ductility. 

\blue{More generally, such control of TB interfacial phases could impact other properties beyond those discussed specifically for Ti above.  For example, the stabilization of complexions that can give rise to solid-state wetting transitions such as those demonstrated by the present calculations can provide pathways for phase transformations with lower nucleation barriers.  Further, the formation of thick complexions like those found here are expected to have consequences for twin nucleation and growth, and also for TB/dislocation interactions that could have important consequences on strength and work hardening. }

\blue{In efforts aimed at control of interfacial phases,} the present work illustrates how the relative stability of competing complexions can be influenced by the free energy difference between stable ($\alpha$ in this case) and metastable polymorph (BCO in this case) phases. Such free energy differences can be tuned through alloying with solute species, in ways that can be predicted through \textit{ab initio} thermodynamics methods or, where appropriate databases are available, computational thermodynamics methods such as CALPHAD\cite{lukas2007}. Hence, for the class of interfacial complexions described in this work, there exists a unique opportunity to guide the design of the interface phases in much the same way that the community has used computational thermodynamics methods to guide the design of microstructures with desired phase combinations and transformations that can be accessed through thermomechanical processing.  The present work thus provides a connection between bulk and interfacial phase diagrams that we expect can be exploited widely in materials design.

\section*{Methods}
\label{sec:methods}

\subsection{Computational methods:}
\label{sec:methods_comp}

DFT calculations are performed with VASP \cite{kresse1996} using the projector augmented wave (PAW) method \cite{kresse1999} within the generalized gradient approximation (GGA) exchange-correlation functional as parameterized by Perdew and Wang \cite{perdew1992} including the 3$p$ electrons as valence. A cutoff energy of $500 \, \text{eV}$ was used for the plane-wave basis set, and the Brillouin zone was sampled using the Methfessel-Paxton method \cite{methfessel1989} (ISMEAR=1 in VASP), with a smearing of $\sigma=0.15 \, \text{eV}$. For the twin boundary supercells, we used a k-point mesh of $6 \times 9 \times 1$, and $4 \times 6 \times 1$ for the calculations presented as a function of strain. The twin boundaries were studied using two different supercell geometries. \blue{In the first setup (i), which was used for all of the reported energies and excess quantities at zero applied strain}, we used a double-TB supercell, i.e., using two twin boundaries along the $z$-direction normal to the interface, to maintain periodicity in this direction with $\sim 38.2 \, \AA$ distance between twin boundaries. \blue{This setup contained $168$ atoms and hereafter is referred to as i.168.} The periodic length along $y=\langle 01\bar{1}0 \rangle$ (the direction parallel to the tilt axis) is $5.086 \, \AA$ and $7.489 \, \AA$ for $x=\langle 4\bar{2}\bar{2}3 \rangle$ (cf., \figext{all_boundaries}). For calculations using this supercell geometry, electronic structure self-consistency loops were terminated at an energy tolerance of $10^{-5} \, \text{eV}$, and in the structural relaxations we fixed the $x$ and $y$ dimensions parallel to the interface, adjusted the $z$ dimension to ensure zero normal stress, and relaxed all atomic positions to a force tolerance of $10 \, \text{meV}\AA^{-1}$. These settings were found to lead to convergence of total energies to approximately $1 \, \text{meV}/\text{atom}$ and TB energies to approximately $0.01 \, \text{J} \, \text{m}^{-2}$. In addition to the calculations for the \tb TB described above, we performed additional calculations for $\{11\bar{2}2\}$, $\{10\bar{1}2\}$, $\{11\bar{2}1\}$, and $\{10\bar{1}1\}$ TBs, using the same approach, with supercells characterized by a periodicity along the $x$ direction of $ 5.498\, \AA$, $6.890 \, \AA$, $ 9.749 \, \AA$, and $10.597 \, \AA$, respectively, and k-point density with the same density as for \tb, corresponding to $8 \times 9 \times 1$, $6 \times 16\times 1$, $4 \times 9 \times 1$, $4 \times 16 \times 1$ meshes, respectively (see \figext{all_boundaries}). 

\begin{figure}[h!]
 \begin{center}
   \includegraphics[scale=0.44]{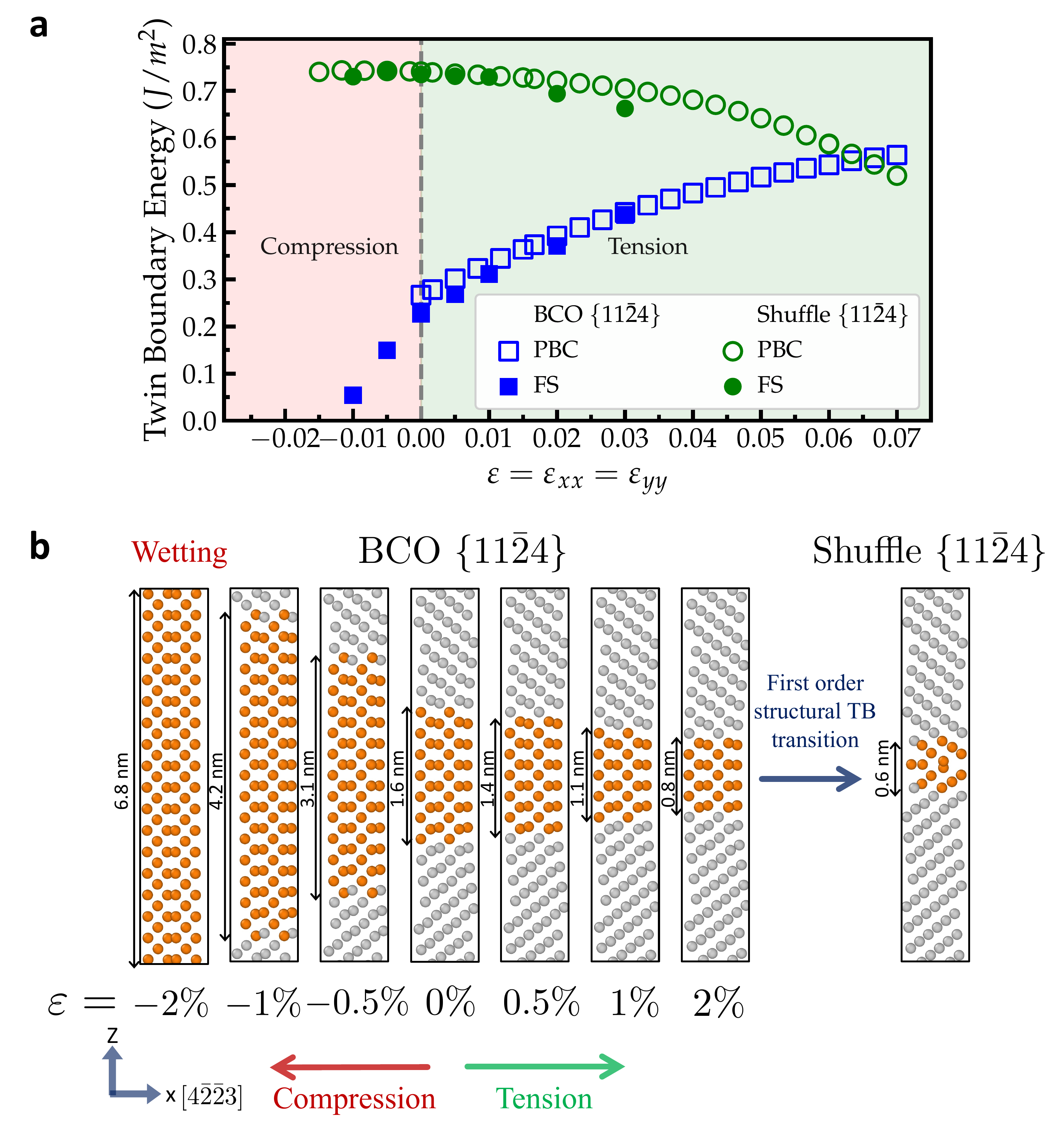}
\caption{Calculated effects of biaxial strain on the \tb twin boundary energy and structure. \textbf{a,} Twin boundary (TB) energy as a function of biaxial strain ($\varepsilon_{\text{xx}}=\varepsilon_{\text{yy}}=\varepsilon $) for the BCO (blue squares) and shuffle (green circles) structures of the \tb TB. \blue{Filled and unfilled symbols correspond to the free-surfaces (FS) and periodic boundary conditions (PBC), respectively (see Methods section).} \textbf{b,} shows the structural changes associated with applied biaxial compressive and tensile strains for the \tbBCO TB \blue{using FS boundary conditions,} illustrating that the width ($w$) of this interface grows (shrinks) with increasing compressive (tensile) strain. The TB supercell is found to undergo a complete transformation to the BCO structure under the compressive biaxial strain of $\varepsilon \approx -1.4 \, \%$ while it transforms to an alternative state with a narrower interfacial thickness through a first-order structural phase transition at the biaxial tensile strain of $\varepsilon = 6.4 \, \%$. \blue{The interfacial width and the color coding} of atoms is based on the CNA \cite{honeycutt1987} with gray and orange representing the HCP and non-HCP (BCO) interfacial regions, respectively. } 
  \label{fig:transitions}
 \end{center}
 \end{figure}

\blue{A second type of supercell geometry (ii), involving the use of free-surface boundary conditions, was used for the calculations of strain-dependent interface structures presented in \fig{transitions}b. They were also used to provide a basis for comparison of  effects of boundary conditions on the TB energies calculated from the periodic supercells (cf., \fig{transitions}a and \tabext{excess}). The supercells with free surfaces had the dimensions of $7.490 \, \times 5.086 \, \times 110 \, \AA$ along $x=\langle 4\bar2\bar23 \rangle$, $y=\langle 01\bar10 \rangle$ and $(11\bar24)$ normal direction, respectively.  A vacuum layer with thickness of $30 \, \AA$ was imposed between the surfaces. The rest of the calculation settings (including convergence criteria and k-point sampling) were the same as given above for the periodic supercell calculations. The interface energy of the single TB at each strain level is calculated by subtracting the total energy from the reference bulk cell with the same number of atoms divided by the TB area \blue{(cf., \tabext{excess})}. Two different system sizes with this setup, both with the same dimensions parallel to the TB plane, but different slab thicknesses in the direction normal to the TB were incorporated: one simulation cell contained $174$ atoms (ii.174), and featured a separation between twin and surfaces of $\sim 39.2 \,\AA$, and the other was $40\,\%$ thicker, with $242$ atoms (ii.242), and a separation between twin and free surfaces of $54.7 \,\AA$. } 

\blue{In all, three different simulation cells have been used in the calculations (i.168), (ii.174) and (ii.242). For zero strain, all of the simulation cells, (i.168), (ii.174) and (ii.242) give rise to TB energies that agree to within $15 \, \text{mJ} \, \text{m}^{-2}$.  The presence of the BCO complexion in the relaxed calculations is robust across all three simulation cells and system sizes, and the thickness of this interface differs by at most one atomic layer between the setup (i.168) and (ii.174) and by less than $1\,\AA$ between the setup (ii.174) and (ii.242) for zero applied strain. Additional calculations using setup (ii.174) were performed under applied compressive and tensile strains ranging between $-2\,\%$ to $3\,\%$, to investigate more fully the effect of the boundary conditions (cf., \fig{transitions}). For the tensile strains, where the interface width stayed narrow, results for thickness and interface energy showed very good agreement (to within $6 \,\AA$ and $30 \, \text{mJ} \, \text{m}^{-2}$, respectively), between (i.168) and (ii.174) supercells, suggesting that the different boundary conditions had minor effect on the calculated results.  Under compressive strains, where the interface thicknesses grew quite large, the free surface boundary conditions were able to accommodate larger widths and gave rise to lower twin-boundary energies.  For these compressive strains the free-surface boundary conditions were viewed to be more reliable and were featured in \fig{transitions}. Overall, these calculations establish that the key computational results used to reach the main conclusions in the paper are robust relative to the boundary conditions and system size.}

The grand-canonical TB evolutionary search is performed using the USPEX code \cite{oganov2006, lyakhov2013}. The details of this method are described in the work by Zhu \et \cite{zhu2018}. In this approach, various mutation operations including the displacement of atoms, insertion and removal of atoms from the TB and sampling of larger-area TB reconstructions are performed to predict the low-energy configurations. \blue{The setup for USPEX is made in a way that we start with the supercell with the TB and apply relative displacements of the two crystals on either side of the interface. In addition, we perform addition and removal of atoms to a sub-region of the structure near the interface, allowing atoms in this region to be relaxed while the rest of the atoms in the bulk regions away from the interface remain constrained. The morphology and the thickness of the interface is found to be insensitive to these constraints. The robustness of these results is validated by two analyses: (i) the obtained structures from the USPEX search have been further relaxed by adding $40 \, \AA$ of vacuum layer on the free surfaces and removing all constrains allowing all the atom positions to relax. No changes in the TB excess energies and interfacial width have been identified in this process; (ii) the same USPEX calculations were also performed around a wider sampling range of $25\,\AA$ subregion around the interface  (compared to the $15\,\AA$ region used in the original search) which is allowed to relax during the search process while keeping all other simulation parameters and inputs the same. Both of these analyses validated the robustness of the results in identifying the identical minimum-energy structures with the same interfacial thickness and TB energy.} The USPEX analysis is performed with classical potential calculations at $0\,$K temperature using the LAMMPS \cite{plimpton1995} software. The modified embedded atom method (MEAM) potential of Hennig \et \cite{hennig2008} was used for evolutionary search calculations. This potential is able to capture the TB excess energy of the ground-state and metastable structures of \tb within $6 \,\, \text{mJ} \, \text{m}^{-2}$ accuracy. Relaxation was terminated when the norm of the forces on all atoms is less than $10^{-4}\,\text{eV}\,{\AA}^{-1}$.

Calculations of the MEP for the $\text{BCO} \rightarrow \omega$ transformation were performed using the $36$-atom supercell taken out from the non-HCP deformed region at the interface (cf., \fig{NEB}). The solid-state nudged-elastic-band (SS-NEB) method \cite{sheppard2012} , as implemented in VASP \cite{henkelman2015}, was used to identify the MEP and associated transition states. The SS-NEB method enables both atomic and cell volume relaxations throughout the MEP search. Calculations were performed using $8$ intermediate images starting from the BCO cell allowing both atomic and cell volume to relax to fully retrieve the $\omega$ phase. We first turned off climbing to verify the transition states and then turned it on to obtain the accurate barrier energy. Then, we fully relaxed the local enthalpy minimum (BCO phase) to verify its stability. The force criterion for SS-NEB convergence was set to be $10 \, \text{meV}\AA^{-1}$. For these calculations, we employed k-point sampling of $6 \times 9 \times 2$ and used the same exchange-correlation potential and plane-wave cutoff as listed above for the TB calculations.

Interfacial excess quantities corresponding to the common TBs in Ti were calculated by DFT as follows. For each structure corresponding to the local minimum of energy, as well as the metastable structure identified for \tb, the excess twin boundary energy ($\gamma$), excess volume per area ($[V]_\text{N}$), and two components of TB stress, $\tau_{xx}$ and $\tau_{yy}$, in the directions perpendicular and parallel to the tilt axis were calculated. These quantities were evaluated from DFT calculations following the equations below: 

\begin{equation}
\begin{aligned}
\gamma= \frac{E_{2\text{TB}}-N \,E_{\text{bulk}}}{2A} ,  \\
[V]_\text{N}= \frac{V_{2\text{TB}}-N \, \Omega_{\text{bulk}}}{2A}
\end{aligned}
\label{eqn:GBE}
\end{equation}

\noindent where $E_{2\text{TB}}$, $E_{\text{bulk}}$, $V_{2\text{TB}}$, $ \Omega_{\text{bulk}}$, $N$ and $A$ are energy of double twin boundary cell, bulk energy per atom for HCP Ti calculated from the same size cell, volume of double twin boundary cell, bulk volume per atom calculated from the same size cell, number of atoms in the double twin cells and TB area, respectively. The factor of two in the denominators of \eqn{GBE} is due to the presence of two interfaces within the periodic supercells, which were used for the calculations of all of the excess-quantity values reported in this manuscript. When defining the TB stress tensor from DFT, we subtracted from the stress of double twin boundary cell constructed from the relaxed lattice dimensions in bulk the stress of the perfect same size (and k-point) supercell. Although this last contribution should be zero in theory for a relaxed box and atomic positions, a small residual stress exists in the perfect crystal because of finite numerical convergence with respect to the plane-wave basis and k-point sampling.

\blue{DFT calculations of the energetics of oxygen solutes in the $\alpha$, BCO, and $\omega$ bulk phases as well as the BCO phase under the strain state found in the TB interface were performed using $96$-atom supercells. The bulk supercells are constructed with lattice vectors parallel to $\text{x}: [2\bar1\bar10]$, $\text{y}: [01\bar10]$, and $\text{z}:[0001]$ for $\alpha$, $\text{x}: [100]$, $\text{y}: [010]$, and $\text{z}:[001]$ for BCO, and $\text{x}: [10\bar10]$, $\text{y}: [\bar12\bar10]$, and $\text{z}:[0001]$ for the $\omega$ phase, see \figsup{oxygen_bulk} for reference. A single oxygen atom is used in supercells with the dimensions $6 \,\times 2 \, \times 2$, $2 \,\times 2 \, \times 2$, and $2 \,\times 2 \, \times 4$ along the lattice directions of the orthogonal bulk $\alpha$, BCO, and $\omega$ cells, respectively with a $2 \times 4 \times 5$ k-point mesh. For the calculation of interaction energies between oxygen and the \tbBCO TB, we made use of the (i.168) supercell and the same k-point mesh, as described above. The convergence criterion and the rest of the DFT settings for the calculations of oxygen energetics in the bulk and at the TB interface are the same as used for the TB relaxations.  }

\subsection{Experimental methods:}
\label{sec:methods_exp}

The Ti-$0.3$ wt.\%O alloy was provided by TIMET, UK. The initial material was argon arc melted and then forged into bars at $1125^\circ$C. The bars were then rolled at $900^\circ$C and annealed at $800^\circ$C for $1$ hour to establish an equiaxed microstructure. The average grain size was determined to be $\sim 60 \; \mu \text{m}$ by Electron backscatter diffraction (EBSD). 

The as-annealed materials were then subjected to cryogenic temperature \blue{($\sim 77 \, \text{K}$)} tensile tests to generate deformation twinning. After the tensile tests, areas near the fracture surface were first mechanically polished and then electro-polished in a solution of $6 \, \%$ perchloric acid and $94 \, \%$ methanol at $-40^\circ$C. EBSD characterizations were carried out at the polished surface using a FEI Strata $235$ scanning electron microscope equipped with a EBSD detector. The \tb twins were identified and marked using the TSL-OIM software. 

Samples for subsequent STEM analysis were made by \blue{focused-ion beam (FIB)}-liftout from an identified \tb twin boundary using a FEI Helios G4 FIB-SEM system. The HAADF-HRSTEM imaging of the twin boundary was conducted on the double-corrected TEAM I microscope (operated at $300 \, \text{kV}$) at the National Center for Electron Microscopy (NCEM), Lawrence Berkeley National Laboratory. \blue{To evaluate the long-range uniformity of the twin boundary phase, additional select area diffraction patterns and  nano beam electron diffraction datasets were also collected. The analysis of the nano beam electron diffraction data was conducted using the py4DSTEM package \cite{savitzky2021}.} 

STEM image simulations were conducted with the DFT-generated atomic configurations using the Prismatic software package for STEM simulation \cite{ophus2017, pryor2017}. The correction parameters (C1, C3 and C5) were set according to the experimental setup on the TEAM I microscope. The thermal effect was omitted to facilitate the image interpretation.

\section*{Acknowledgments}
\label{sec:ackn}
The authors would like to thank Professor Daryl C. Chrzan for valuable discussions. This research was supported by the Office of Naval Research under Grant no. N00014-19-1-2376. This work was partially performed under the auspices of the U.S. Department of Energy (DOE) by the Lawrence Livermore National Laboratory (LLNL) under Contract No. DE-AC52-07NA27344. E.C. acknowledges support from a National Science Foundation Graduate Research Fellowship Program under Grant No. DGE-1752814, and T.F. acknowledges the funding by the Laboratory Directed Research and Development Program at LLNL under Project Tracking Code number 19-ERD-026. Work at the Molecular Foundry was supported by the Office of Science, Office of Basic Energy Sciences, of the U.S. Department of Energy under contract no. DE-AC02-05CH11231. This work made use of computational resources provided by the Extreme Science and Engineering Discovery Environment (XSEDE), which is supported by National Science Foundation grant number ACI-1548562, LLNL Institutional Computing facilities, and the Savio computational cluster resource provided by the Berkeley Research Computing program at the University of California, Berkeley (supported by the UC Berkeley Chancellor, Vice Chancellor for Research, and Chief Information Officer).

\section*{Author Contribution}
\label{sec:AC}

M.S.H, M.A, R.Z, Y.C and A.M.M designed the simulations and experiments. M.S.H performed all the DFT calculations, and led the analysis with input from M.A., D.L.O., and T.F.  Y.C and R.Z conducted and analyzed the experiments, and performed image simulations and analysis with input from M.S.H, and A.M.M.  E.C., and T.F. performed the evolutionary search simulations, and worked with M.S.H on their analysis.  All authors contributed to the writing of the manuscript.

\section*{Data availability}
\label{sec:data}
The data that support the findings of this study are available from the corresponding author upon reasonable request.

\section*{Competing interests}
\label{sec:CI}

The authors declare that they have no competing interests.

\section*{}

\pagebreak
\newpage
\clearpage

\section*{References}
\bibliographystyle{sh}

%
%

\pagebreak
\newpage
\clearpage


 \section*{}

\pagebreak
\newpage
\clearpage

\setcounter{table}{0}
\renewcommand{\tablename}{Extended Data Tab.}%

 \setcounter{figure}{0}
\renewcommand{\figurename}{Extended Data Fig.}%

 \begin{figure}[h!]
 \begin{center}
   \includegraphics[width=\textwidth]{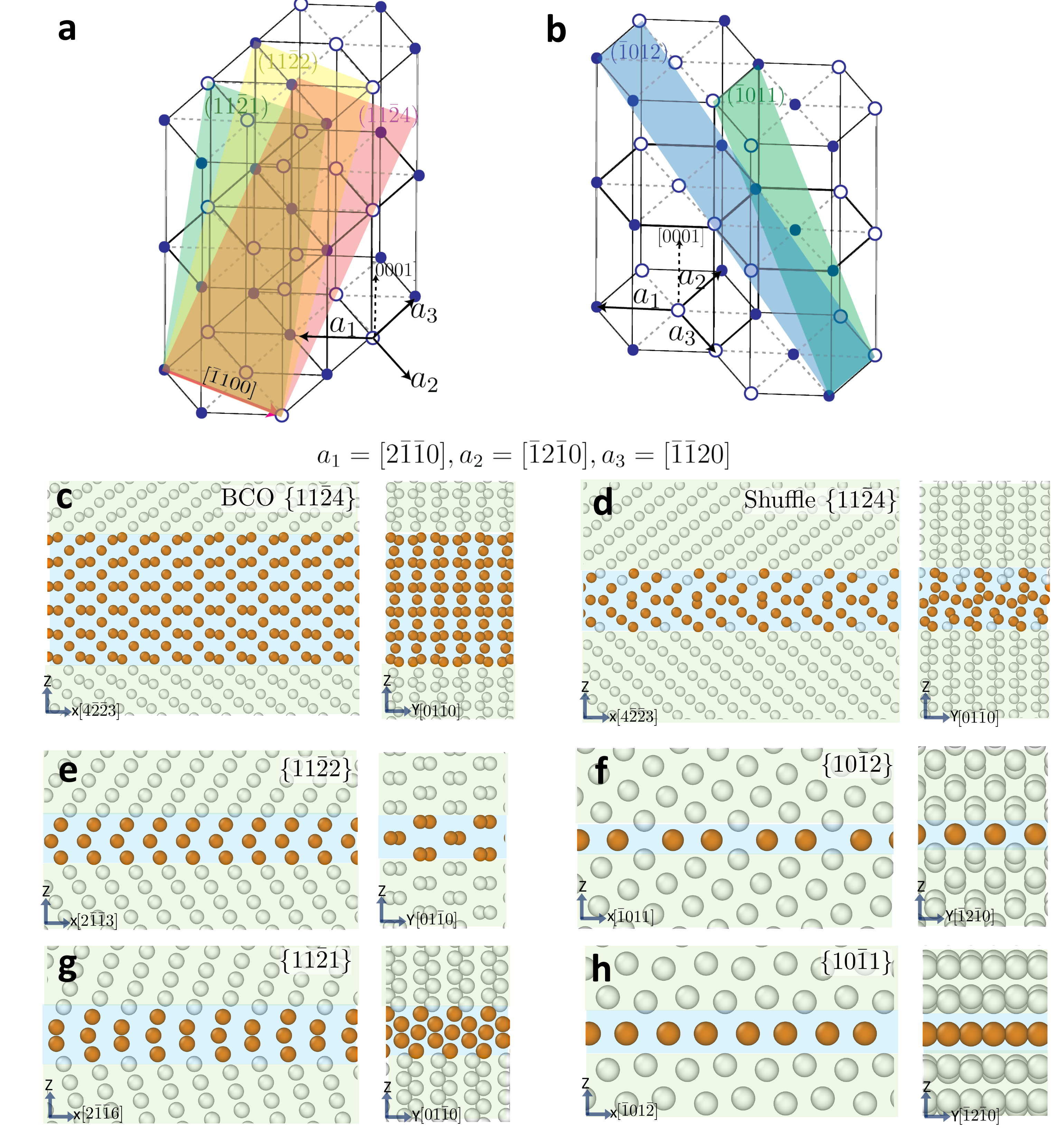}
\caption{Crystallographic representations of $\langle 1\bar100\rangle$ \textbf{(a)} and $\langle \bar12\bar10\rangle $ \textbf{(b)} twin boundary tilt angles in an HCP crystal lattice. Atoms on two distinct $(\bar12\bar10)$ \textbf{(a)} and $(\bar1\bar120)$ \textbf{(b)} planes are distinguished by the solid and empty circles. DFT relaxed TB structure of \textbf{c} \tbBCO,  \textbf{d} \tbshufff,  \textbf{e} $\{11\bar22\}$,  \textbf{f}  $\{10\bar12\}$,  \textbf{g} $\{11\bar21\}$, and  \textbf{h} $\{10\bar11\}$ are represented. Color coding is the CNA \cite{honeycutt1987} with gray representing HCP and orange showing non-HCP regions.}
  \label{fig:all_boundaries}
 \end{center}
 \end{figure}

 \begin{figure}[h!]
 \begin{center}
  \includegraphics[scale=0.6]{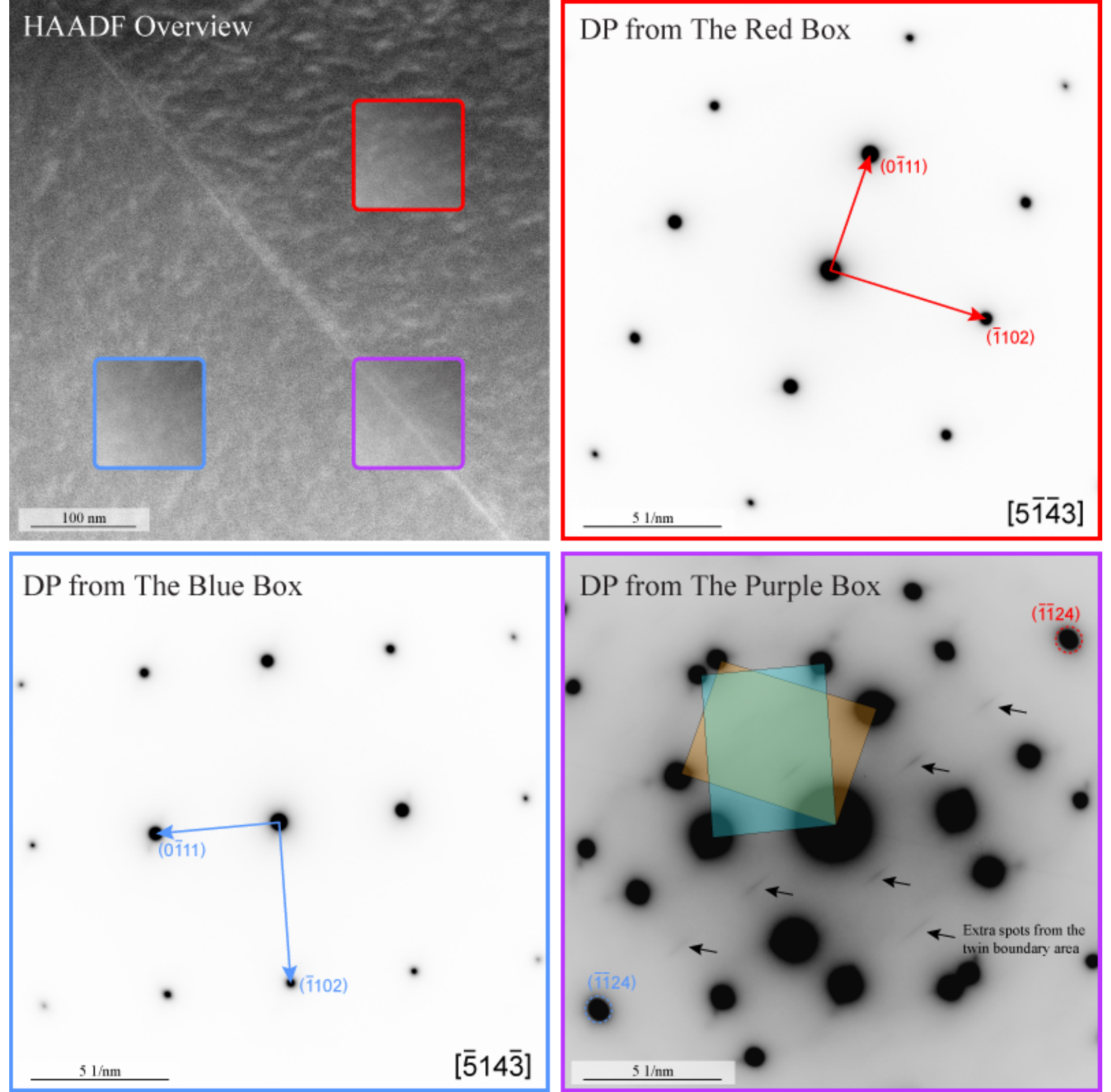}
\caption{Measured selected-area diffraction (SAD) patterns in the vicinity of a $\{11\bar24 \}$ twin boundary in a cryo-deformed Ti-$0.3$ wt.\%O  alloy. The high-angle angular dark field (HAADF) image shows the representative area around the twin boundary and indicates the aperture positions for the three SAD patterns. The red- and the blue-framed diffraction patterns (DP) illustrate the orientations of the matrix and the twin, respectively. The purple-framed DP demonstrates the mirror relation of the diffraction peaks from both sides of the twin boundary, with noticeable streak-shaped extra spots from the BCO phase (marked by the black arrows). The g vectors of the twinning plane, $(\bar1\bar124)$, from the matrix and the twin are marked by the dashed red and the blue circles, respectively. The contrast of the DPs are inverted for better visibility (dark is high electron intensity).}
  \label{fig:FFT}
 \end{center}
 \end{figure}

 \begin{figure}[h!]
 \begin{center}
   \includegraphics[width=\textwidth]{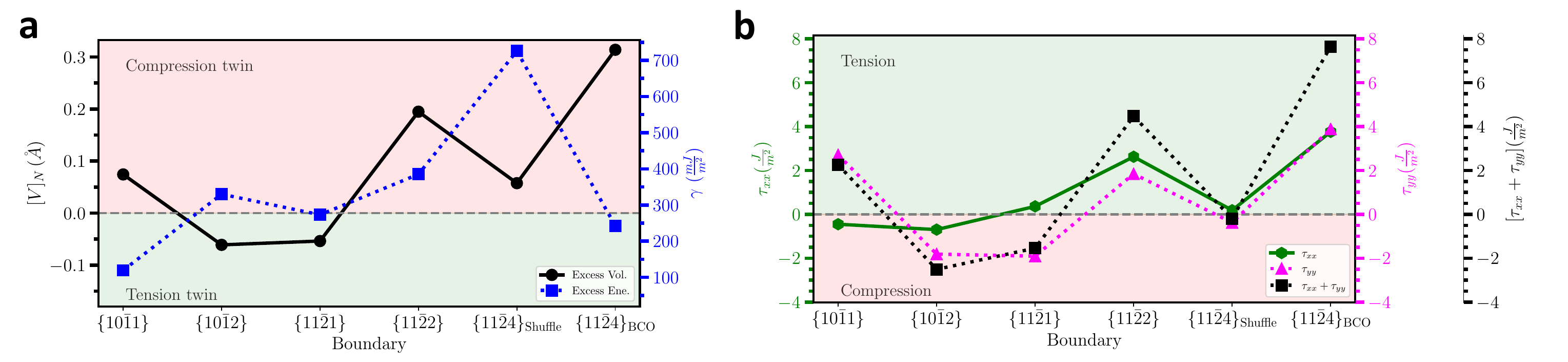}
\caption{ Excess thermodynamic properties of twin boundaries in titanium. \textbf{a,} DFT calculated excess volume ($[V]_N$) and excess TB energy  ($\gamma$) for different TBs in Ti;  \textbf{b,} TB interfacial stress normal ($\tau_{xx}$) and parallel ($\tau_{yy}$) to the tilt axis, respectively.  }
  \label{fig:energies}
 \end{center}
 \end{figure}

\clearpage

\begin{table}[ht]
\caption{ Excess thermodynamic properties of twin boundaries in titanium. DFT calculated excess volume per area ($[V]_{N} (\AA)$), excess TB energy ($\gamma [\frac{mJ}{m^2}]$), TB interfacial stress normal ($\tau_{xx} [\frac{J}{m^2}]$) and parallel ($\tau_{yy} [\frac{J}{m^2}]$) to the tilt axis for different TBs in Ti \blue{using periodic boundary conditions (see Methods). Values in the parentheses show the excess quantities for the \tb TB using free-surface boundary conditions. }  }

\label{tab:excess}
\centering
\begin{tabular}{lllll}
\hline
TB                              & $[V]_{N} (\AA)$                                         & $\gamma (\frac{mJ}{m^2})$                                   & $\tau_{xx} (\frac{J}{m^2})$                              & $\tau_{yy} (\frac{J}{m^2})$                               \\ \hline
$\{10\bar11\}$                  & 0.074                                                   & 119.391                                                     & -0.450                                                   & 2.709                                                     \\
$\{10\bar12\}$                  & -0.061                                                  & 329.972                                                     & -0.696                                                   & -1.813                                                    \\
$\{11\bar21\}$                  & -0.054                                                  & 273.050                                                     & 0.364                                                    & -1.906                                                    \\
$\{11\bar22\}$                  & 0.195                                                   & 384.957                                                     & 2.626                                                    & 1.847                                                     \\
$\{11\bar24\}~(\text{Shuffle})$ & \begin{tabular}[c]{@{}l@{}}0.057\\ (0.071)\end{tabular} & \begin{tabular}[c]{@{}l@{}}725.812\\ (734.729)\end{tabular} & \begin{tabular}[c]{@{}l@{}}0.186\\ (0.0620)\end{tabular} & \begin{tabular}[c]{@{}l@{}}-0.376\\ (-0.213)\end{tabular} \\
$\{11\bar24\}~(\text{BCO})$     & \begin{tabular}[c]{@{}l@{}}0.314\\ (0.335)\end{tabular} & \begin{tabular}[c]{@{}l@{}}242.582\\ (226.808)\end{tabular} & \begin{tabular}[c]{@{}l@{}}3.762\\ (2.830)\end{tabular}  & \begin{tabular}[c]{@{}l@{}}3.888\\ (3.001)\end{tabular}   \\ \hline
\end{tabular}
\end{table}

%% file: Twin_Boundary_Structural_Phase_Transitions_in_Elemental_Titanium_sup_rev.tex
\pagebreak
\newpage
\clearpage

\title{ {\large Supplementary Information for}\\[0.5cm] { \Large \textbf{Twin-Boundary Structural Phase Transitions in Elemental Titanium}}}\\
\author{Mohammad S. Hooshmand$^{1}$, Ruopeng Zhang$^{1,2}$, Yan Chong$^{1,2}$, Enze Chen$^{1}$, Timofey Frolov$^{4}$, David L. Olmsted$^{1}$,  Andrew M. Minor$^{1,2,3}$ \& Mark Asta$^{1,3*}$  }
\date{\\[0.5cm]
    \normalsize{$^1$Department of Materials Science and Engineering, University of California, Berkeley, CA 94720, USA. \\ $^2$ National Center for Electron Microscopy, Molecular Foundry, Lawrence Berkeley National Laboratory, Berkeley, CA 94720, USA. \\ $^3$ Materials Sciences Division, Lawrence Berkeley National Laboratory, Berkeley, CA 94720, USA. \\$^4$  Lawrence Livermore National Laboratory, Livermore, CA 94550, USA. \\ $^{*}$  Email: \href{mailto: mdasta@berkeley.edu}{mdasta@berkeley.edu} }}

\pagebreak
\newpage
\clearpage

\normalsize

\beginsupplement

\section{Further experimental evidence on the \tb twin boundary structure }
\label{secsup:more_exp}

\begin{figure}[h!]
 \begin{center}
   \includegraphics[width=\textwidth]{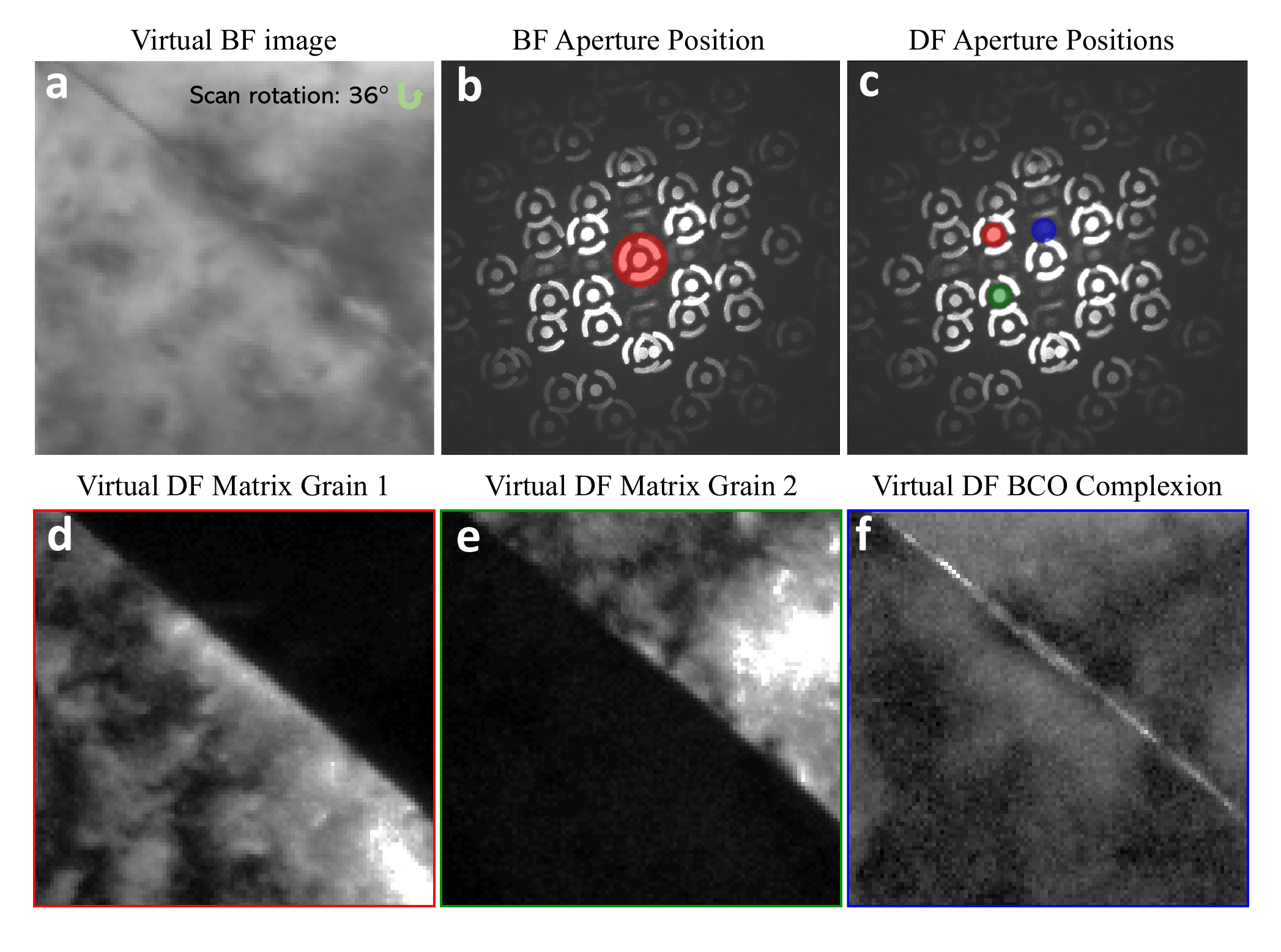}
\caption{Results of nano beam electron diffraction experiments. \textbf{a,} a reconstructed virtual bright-field (BF) image showing an area of intermediate thickness around the twin boundary. \textbf{b,} the maximum value diffraction pattern from the nano beam diffraction dataset. The red circle indicates the size and position of a virtual aperture used to reconstruct the BF image in a. \textbf{c,} the maximum value diffraction pattern from the nano beam diffraction dataset. The red, green and blue circles indicate the size and position of the virtual apertures used to reconstruct the dark-field (DF) images in \textbf{d, e, f,} respectively. \textbf{d-f}, virtual DF images showing the two matrix grains and the twin boundary phase. The contrast from the twin boundary phase is constantly around $1-2 \, \text{nm}$ throughout the entire region of interest. 
  }
  \label{figsup:4DSTEM}
 \end{center}
 \end{figure}

Due to the wedge shape of the focused-ion beam (FIB)-milled sample, there is only a limited region that is feasible for atomic resolution scanning transmission electron microscopy (STEM) imaging on the sample. To further investigate the existence and the uniformity of the twin boundary (TB) phase, a nano beam electron diffraction (4D STEM) experiment was conducted at a region with intermediate thickness. The probe size used in the experiment is $\sim 1 \, \text{nm}$, and a $100 \times 100$ scan (in real space) with a step size of $1 \, \text{nm}$ is collected and analyzed. The results are summarized in \figsup{4DSTEM}. According to the virtual dark-field (DF) image of the TB phase (\figsup{4DSTEM}f), the thickness of the boundary phase is around $1-2 \, \text{nm}$ at a relatively thick region $(> 100 \, \text{nm})$ of the FIBed sample, consistent with the results obtained by high-resolution imaging in the thinner part of the sample. This result confirms that the presence of the body-centered orthorhombic (BCO) complexion for the TB is found in both thick and thin regions of the sample, and its existence is not attributed to surface effects in the thinnest region of the sample.

\begin{figure}[h!]
 \begin{center}
   \includegraphics[width=\textwidth]{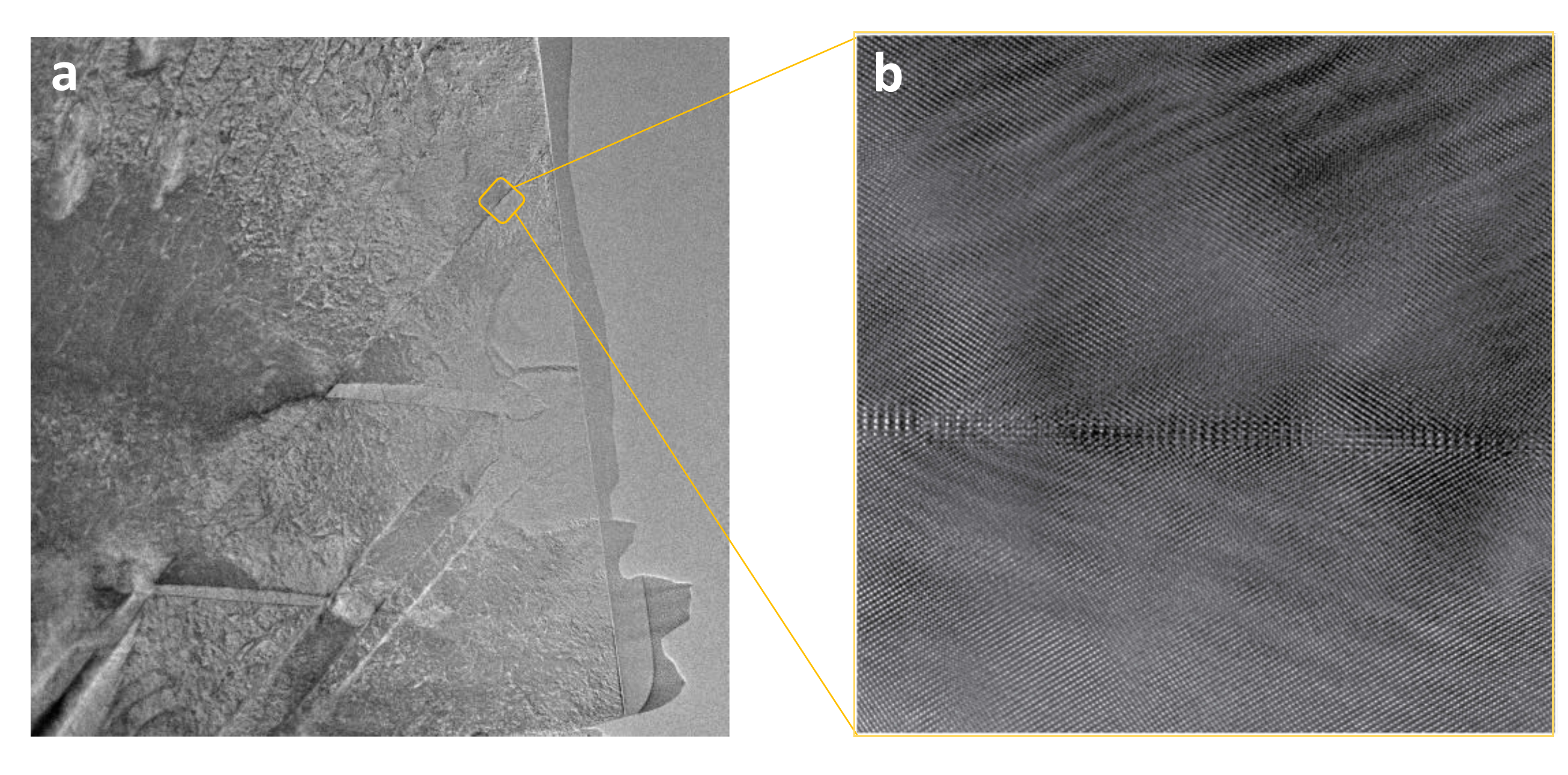}
\caption{ Additional transmission electron microscopy (TEM) observations of a sample containing the \tb twins. \textbf{a,} a low magnification TEM bright-field image showing the extensive deformation in the vicinity of the observed twin. \textbf{b,} a HRTEM image showing the phase contrast caused by defects near the twin boundary.  }
  \label{figsup:HRTEM}
 \end{center}
 \end{figure}

\begin{figure}[h!]
 \begin{center}
   \includegraphics[width=\textwidth]{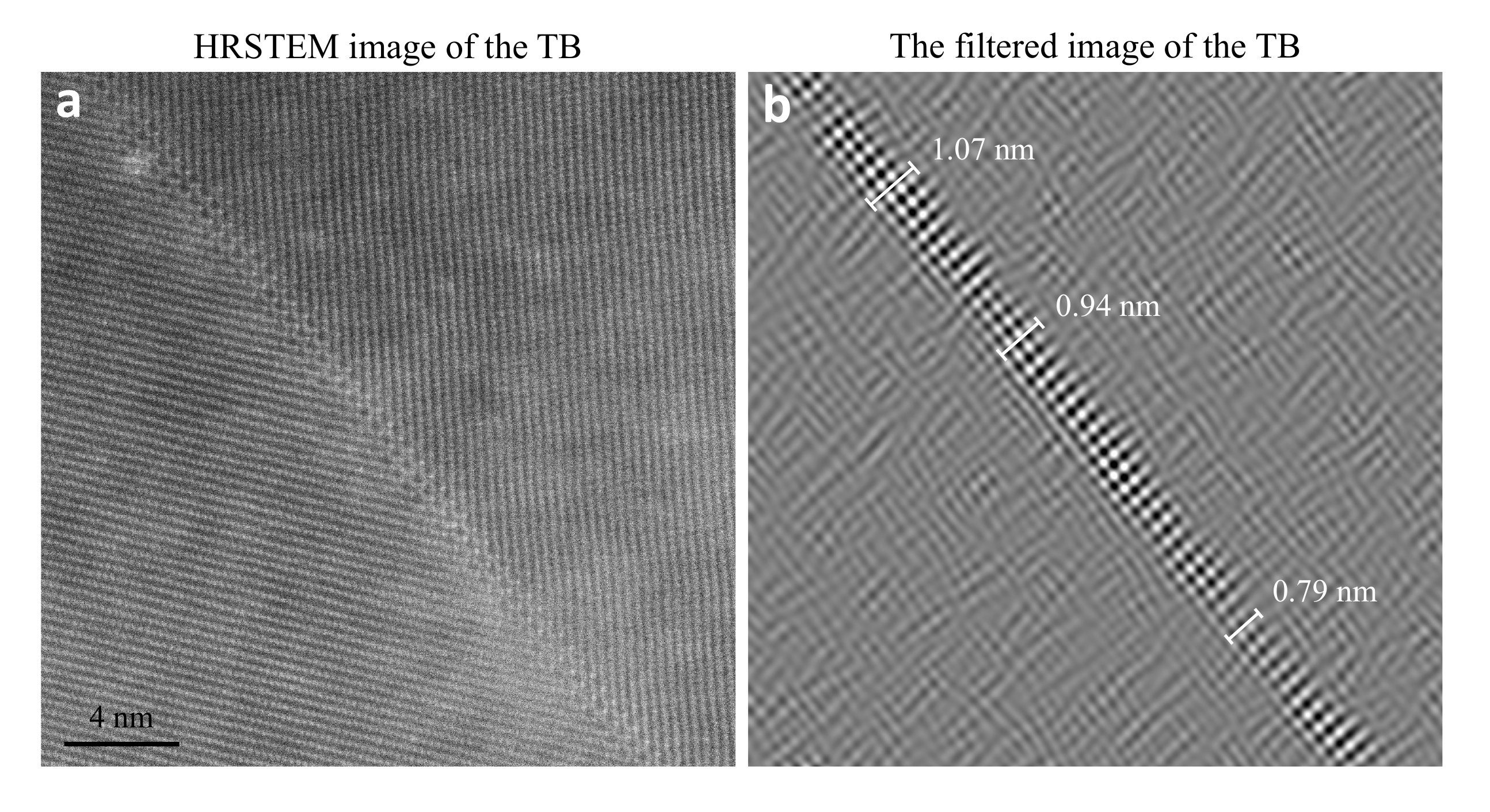}
\caption{Additional HRSTEM observation of a sample containing \tb twins. \textbf{a,} a HRSTEM image showing clear contrast of the BCO interfacial complexion. \textbf{b,} filtered HRSTEM image from the boundary phase Fast Fourier transform (FFT) spots. The fluctuation of the thickness of the boundary complexion is demonstrated and measured. 
  }
  \label{figsup:HRSTEM_width}
 \end{center}
 \end{figure}

From measurement of the local strain across the sample, we find appreciable heterogeneous variations, as expected due to the presence of a high density of defects introduced in the deformation process used to form the TBs (See \figsup{HRTEM}). Similarly, we find that the local interface width of the TB also fluctuates along the length of the interface, as demonstrated in \figsup{HRSTEM_width}. These observations are discussed in the main context, in the context of the disjoining potential formalism, in which local variations in stress should alter the bulk energy difference between the BCO and hexagonal-close-packed (HCP) structures, which is expected to impact the local interfacial width.

\pagebreak
\newpage
\clearpage

\section{Model for the stability of BCO complexion}
\label{secsup:disjoining}

\begin{figure}[h!]
 \begin{center}
   \includegraphics[width=\textwidth]{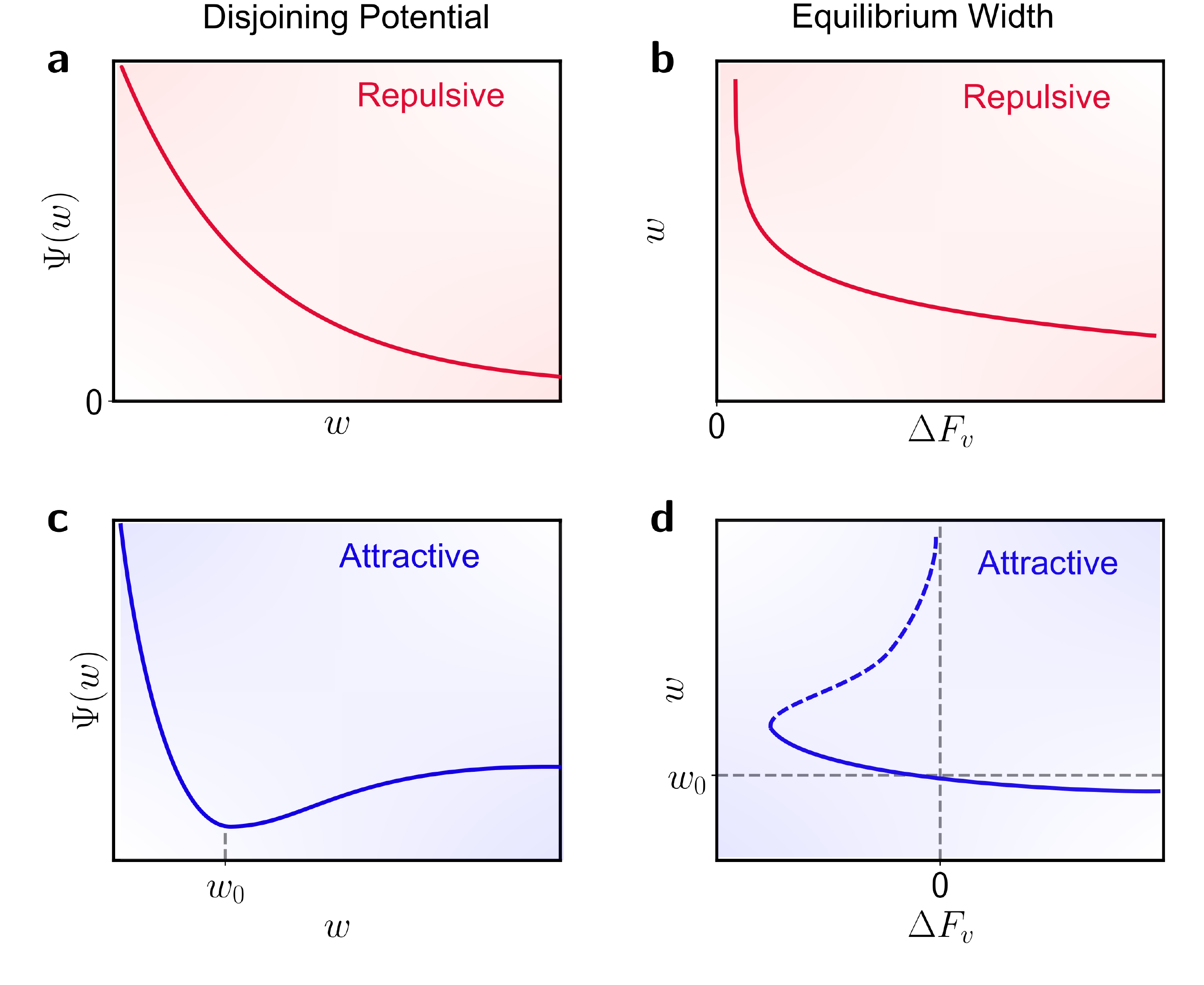}
\caption{Classification of disjoining potential ($\Psi$) into representative repulsive and attractive cases. \textbf{a,} \textbf{c,} Potential shapes as a function of width ($w$). \textbf{b,} \textbf{d,} Schematic behavior of the equilibrium interfacial widths as a function of bulk free energy difference between the two phases ($\Delta F_v$). In \textbf{d}, the solid line corresponds to widths for stable (for $\Delta F_V > 0$) and metastable (for $\Delta F_V < 0$) minima in the interfacial structure, while the dashed line (for $\Delta F_V < 0$) corresponds to a local maximum in the excess energy versus width. Top and bottom panels correspond to the repulsive and attractive potentials, respectively.}
  \label{figsup:wet}
 \end{center}
 \end{figure}

The discovery of the solid-state wetting transition under applied compressive strain suggests a model for the stability of the \tbBCO complexion using concepts from interfacial wetting theory.  In this model we make use of a thermodynamic formalism used for grain-boundary melting phenomena \cite{lipowsky1987, fisher1985}, in which the total excess free energy per area $F(w)$ of a system containing a \tbBCO TB with width $w$ is written as follows: 

\setcounter{equation}{1}
\begin{equation}
F(w) = \Delta F_V \, w + \Psi(w)
\label{eqn:G_psi}
\end{equation}

\noindent where $\Delta F_V = F_V^{\text{BCO}} – F_V^{\text{HCP}}$ represents the free energy difference per volume between the BCO ($F_V^{\text{BCO}}$) and HCP $\alpha$-Ti ($F_V^{\text{HCP}}$) phases, and $\Psi(w)$ is the so-called ``disjoining potential” that represents a width-dependent interfacial free energy. At zero temperature, $\Delta F_V$ is proportional to the energy difference $\Delta E$ between the BCO and HCP. The disjoining potential takes values of $\gamma$ for a hypothetical \tb TB with no BCO wetting phase in the limit of $w \rightarrow 0$ and twice the interfacial free energy for a BCO/HCP heterophase interface in the limit $w \rightarrow \infty$. Two qualitatively different types of behavior for $\Psi(w)$ are typically assumed, which we will refer to as \textit{repulsive} and \textit{attractive}, as illustrated in \figsup{wet}.  

In the repulsive case, $\Psi(w)$ decreases monotonically as a function of $w$ (\figsup{wet}a), while in the attractive case $\Psi(w)$ contains a minimum at a width $w_0$ (\figsup{wet}c).  For repulsive $\Psi(w)$ the wetting transition is continuous in nature, with the interface width diverging as the bulk phase transition is approached from above, i.e., as $\Delta F_V$ approaches zero from the region above the transition where $\Delta F_V > 0$ (\figsup{wet}b).  By contrast, if $\Psi(w)$ is attractive the wetting transition is first-order in character, with a finite-width interface being stable above and below the phase transition (\figsup{wet}d), until a limiting negative value of $\Delta F_V(w)$ is reached that corresponds to an instability point, where the finite-width interface is unstable and the system transforms to (i.e., is fully wetted by) the equilibrium phase.  The behavior shown in \fig{transitions}b is consistent with such first-order wetting behavior, as discussed in further detail below.

\begin{figure}[h!]
 \begin{center}
   \includegraphics[scale=1]{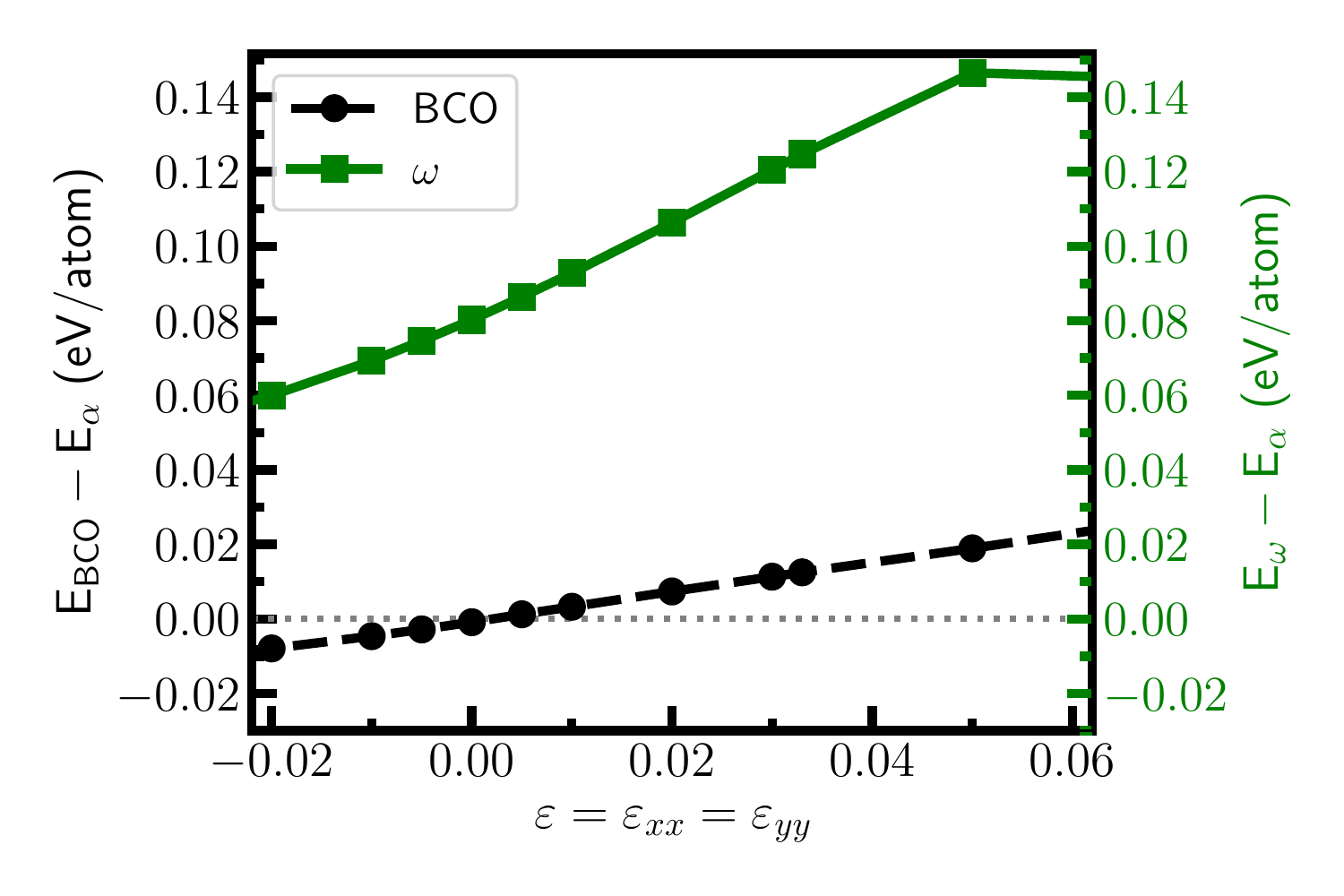}
\caption{\blue{The DFT-GGA calculated bulk energy differences (per atom) between the BCO and HCP (black circles) and the $\omega$ and HCP (blue squares) phases of titanium as a function of applied biaxial strain ($\varepsilon = \varepsilon_{\text{xx}}=\varepsilon_{\text{yy}}$). In these results, BCO and $\omega$ have been strained to the values corresponding to the state of strain imposed by the \tb TB (see main text). In other words, the data point at zero applied strain corresponds to the strain state of the \tb twin boundary in which the BCO phase is under tensile stress imposed by the stress-free bulk HCP structure at the interface. }}
  \label{figsup:DFV}
 \end{center}
 \end{figure}

\blue{For density functional theory (DFT) calculations using the generalized gradient approximation (GGA) exchange-correlation functional (hereafter is referred to as DFT-GGA)} at zero temperature, we note that the strained BCO phase is slightly lower in energy than $\alpha$-Ti (see \tabsup{wet} and \figsup{DFV}) for values of $\varepsilon$ that are zero. \blue{It is also evident from \figsup{DFV} that the strained $\omega$ is slightly higher in energy than strained BCO which explains that the complexion with BCO morphology is favored compared to $\omega$ at the interface of the \tb TB. } As shown in \figsup{DFV}, for increasing compressive strains the BCO phase becomes more stable (i.e., $\Delta F_V$ becomes increasingly negative), while for increasing tension HCP becomes more stable (i.e., $\Delta F_V$ becomes increasingly positive). The fact that a finite width interface is found by DFT at zero and slightly negative values of $\varepsilon$ where the BCO phase is stable suggests that within the framework described above, the \tbBCO interface is characterized by a disjoining potential with an attractive minimum. This can be seen from \figsup{wet}d, where a finite width for $\Delta F_V < 0$ requires an attractive $\Psi(w)$ (\figsup{wet}c). 

This interpretation would then provide an explanation for why the width of the interface calculated by DFT is larger than that observed experimentally. Namely, for this type of wetting transition, the (stable or metastable) equilibrium interface width ($w_{\text{eq}}$) is that which leads to a (global or local) minimum of \eqn{G_psi}. Beyond the transition, where $\Delta F_V > 0$, $w_{\text{eq}}$ will be smaller than $w_0$ since the bulk free energy will favor reduction in the volume of the higher-energy phase.  By contrast, below the transition, where $\Delta F_V < 0$, $w_{\text{eq}} > w_0$ since the $\Delta F_V$ will favor an increase in the volume of the lower-energy bulk phase.  While in the zero-temperature DFT-GGA calculations for pure titanium at $\varepsilon=0$ the strained BCO phase has lower energy than $\alpha$-Ti, i.e., $\Delta F_V < 0 $, we expect the opposite to be true in the room-temperature measurements for $\alpha$ Ti-O alloys, since $\alpha$, with the HCP structure, is the known equilibrium phase. In this case it would follow that $w_{\text{eq}}$ in the experimental measurements would be lower than calculated by DFT, consistent with the results shown in \fig{TEM}.

  \section*{}

\pagebreak
\newpage
\clearpage

 \begin{table}[]
\caption{ Structural parameters and the DFT-GGA calculated thermodynamic quantities.  $\text{L}$, $\text{A}$, $N_\text{BCO}$, $N_{\alpha}$, $\text{V}^{\text{atom}}_\text{BCO}$, $\text{V}^{\text{atom}}_{\alpha}$,   $\text{E}^{\text{atom}}_{\text{BCO}} $, and $\text{E}^{\text{atom}}_{\alpha}$ are the double twin boundary cell height (in $\AA$), twin boundary area (in $\AA^2$), number of atoms corresponding to BCO and $\alpha$-HCP identified in double twin boundary cell (per each interface), volume of BCO and $\alpha$ phases per atom (in $\AA^3$), and calculated energy per atoms (in eV), respectively. The values in the parentheses correspond to the strained BCO under the in-plane lattice constraint imposed by $\alpha$ matrix. }

\label{tabsup:wet}
\begin{adjustbox}{center}

\begin{tabular}{cccccccc}
\hline
$\text{L}(\AA)$ & $\text{A}(\AA^2)$ & $N_\text{BCO}$ & $N_{\alpha}$ & $\text{V}^{\text{atom}}_\text{BCO} (\AA^3)$      & $\text{V}^{\text{atom}}_{\alpha} (\AA^3)$ & $\text{E}^{\text{atom}}_{\text{BCO}} (\frac{\text{eV}}{\text{atom}})$           & $\text{E}^{\text{atom}}_{\alpha} (\frac{\text{eV}}{\text{atom}})$ \\ \hline
77.168          & 38.089            & 34             & 50           & \begin{tabular}[c]{@{}c@{}}17.379\\ (17.603)\end{tabular} & 17.361                                    & \begin{tabular}[c]{@{}c@{}}-7.8049\\ (-7.8029)\end{tabular} & -7.8024                                       \\ \hline
\end{tabular}
\end{adjustbox}
\end{table}

\section{Crystallographic charactristics of the \tb twin boundary structures}
\label{secsup:crystal}

\begin{figure}[h!]
 \begin{center}
  \includegraphics[scale=0.5]{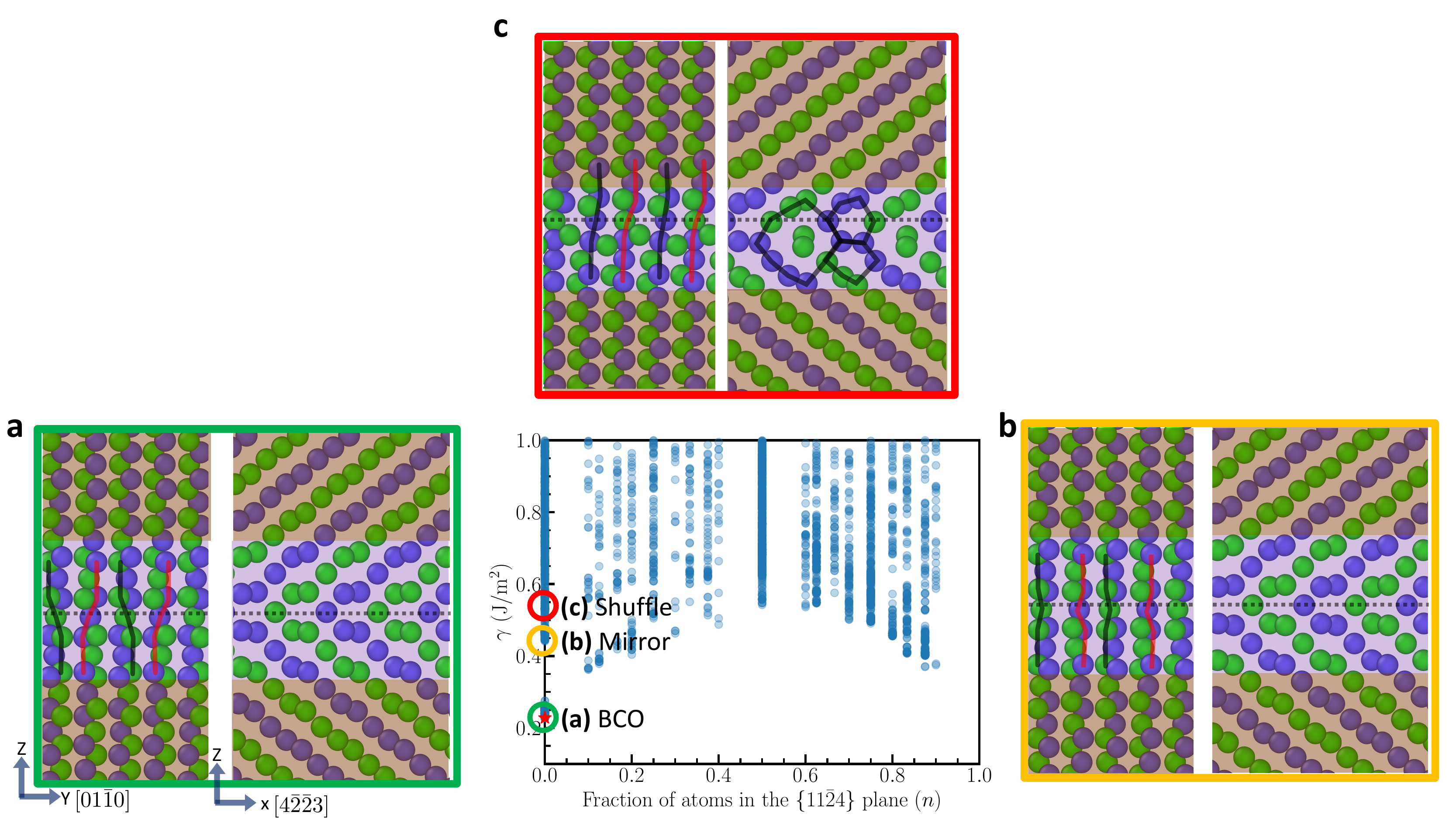}
\caption{ Geometrical interpretation of atomic arrangements at the interface of the \tb TB. The results of the grand-canonical structure search for the \tb TB in Ti from \fig{unique} is repeated in the center. The ground state structure is indicated by the green circle. The yellow circle corresponds to a higher energy state in which the mirror symmetry between the parent and twinned crystal is preserved. The red circle indicates the competing shuffle state with higher excess energy. The atomic structures of these states relaxed using the MEAM interatomic potential are shown in \textbf{a-c}. Color coding of atoms are based on the two atom types per primitive cell of HCP lattice distinguishing every other basal planes. Colored lines are the twin-matrix traces of $\{01\bar10\}$ planes highlighting the structural differences between these structures. Comparison between the atomic configuration of the ground state structure shown in panel \textbf{a} with \textbf{b} highlights that the twinned crystal does not satisfy mirror symmetry.  }
  \label{figsup:uspex}
 \end{center}
 \end{figure}

Twinning operations can either correspond to a reflection with respect to the twinning plane, $K_1$, (Type I) or a rotation of $180 ^\circ$ along the twinning direction $\eta_1$(Type II). A previous study by Minonishi and co-workers \cite{minonishi1982} demonstrated that the minimum energy structure of twinned $\{11\bar21\}$ does not satisfy mirror symmetry. Instead, the alternate basal planes are displaced by $\frac{a\sqrt{3}}{12}$ normal to the plane of shear in the opposite direction which leads to a change in the stacking sequence. For the case of the \tb TB in Ti, we find that both modified embedded atom method (MEAM) and DFT predict a lower twin boundary energy than the mirror reflection and that of corresponding to the $\{11\bar21\}$ atomic structure. The minimum energy atomic structure for the case of \tb TB is achieved by translating consecutive basal planes by $\frac{a\sqrt{3}}{6}$ in the opposite direction along normal to the twinning direction (leading to the \tbBCO in \figsup{uspex}a). Pure mirror symmetry leads to the meta-stable structure of illustrated in \figsup{uspex}b, which corresponds to the morphology with higher energy than the \tbBCO structure (different stacking than BCO along the $c$ direction). The Shuffle structure can be obtained by displacing successive basal planes by $\frac{a\sqrt{3}}{3}$ (\figsup{uspex}c).

\pagebreak
\newpage
\clearpage

\section{Effect of exchange-correlation functionals on the \tb TB structure}
 \label{secsup:exchange}
 
 Here, we study the effect of exchange-correlation (XC) functional on DFT-calculated structure for the \tb TB. Specifically, we compare the results from our DFT-GGA calculations with those using DFT+U (GGA+U here) functional with on-site Hubbard corrections \cite{dudarev1998}. The inclusion of +U has been proposed in previous calculations for Ti \cite{dudarev1998,lutfalla2011,zarkevich2016} to improve the description of the 3$d$ electron states. Specifically, we employ the Dudarev \et formalism \cite{dudarev1998} with $U-J = 2.2 \,\text{eV}$, which has been reported to reproduce the observed energy of reduction in Ti oxides \cite{lutfalla2011}. We start by calculating the relative structural enthalpies of $\alpha$, BCO and $\omega$ phases. \tabsup{DFTU} compares the relaxed lattice constants and relative structural enthalpies of these phases using the GGA and GGA+U functionals. GGA+U predicts the HCP phase as the ground state of Ti at zero temperature and pressure. 
 
\begin{figure}[h!]
 \begin{center}
   \includegraphics[width=\textwidth]{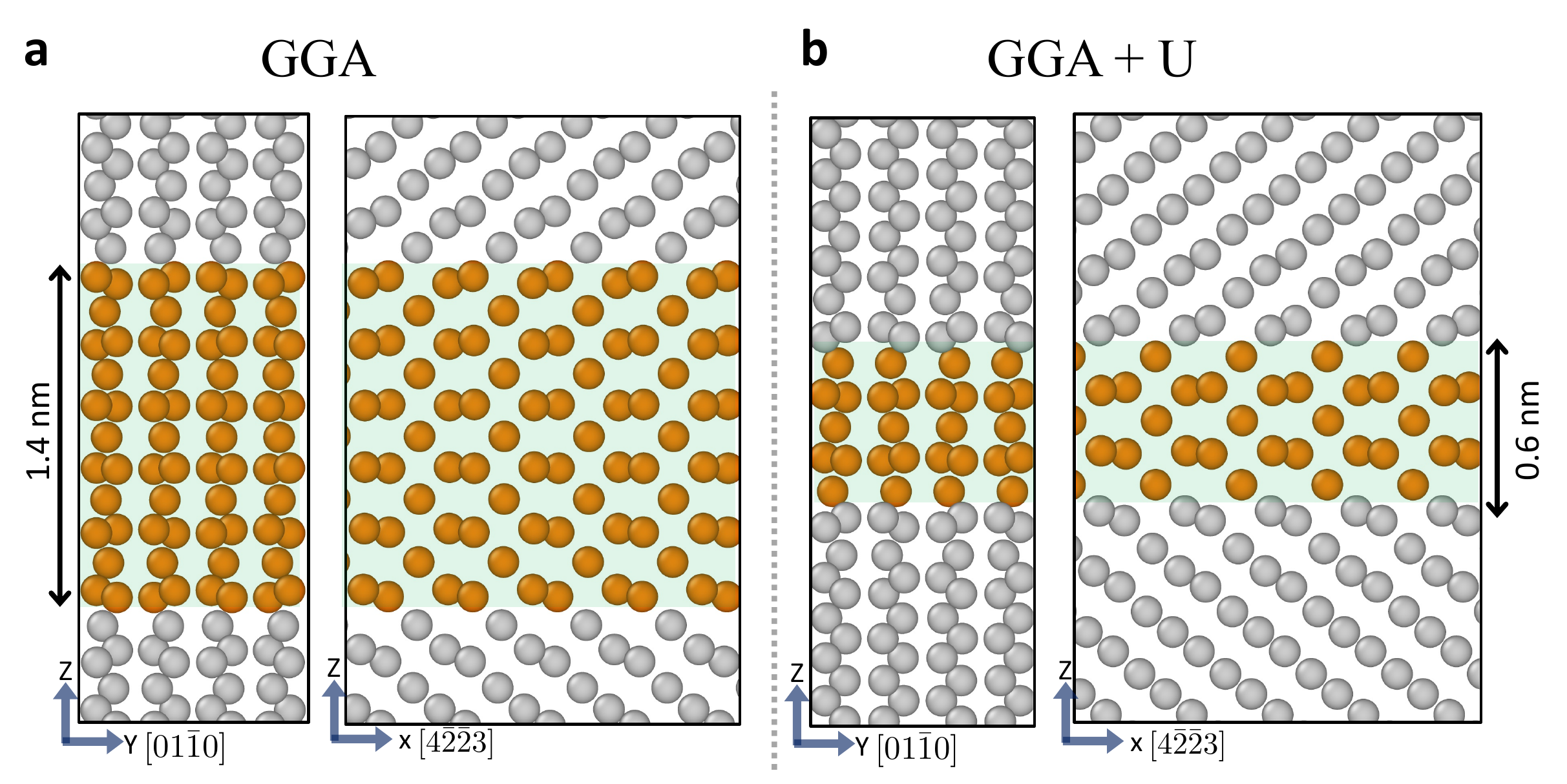}
\caption{ Relaxed structure of the \tbBCO TB using \textbf{a,} GGA and \textbf{b,} GGA+U exchange correlation functionals. BCO structure is formed at the interface using both exchange correlation functionals with a reduced width in GGA+U case. The interfacial width and the color coding of atoms is based on the CNA \cite{honeycutt1987} with gray and orange representing the HCP and non-HCP (BCO) interfacial regions, respectively.     }
  \label{figsup:GGAU_TB}
 \end{center}
 \end{figure}

\begin{figure}[h!]
 \begin{center}
   \includegraphics[width=\textwidth]{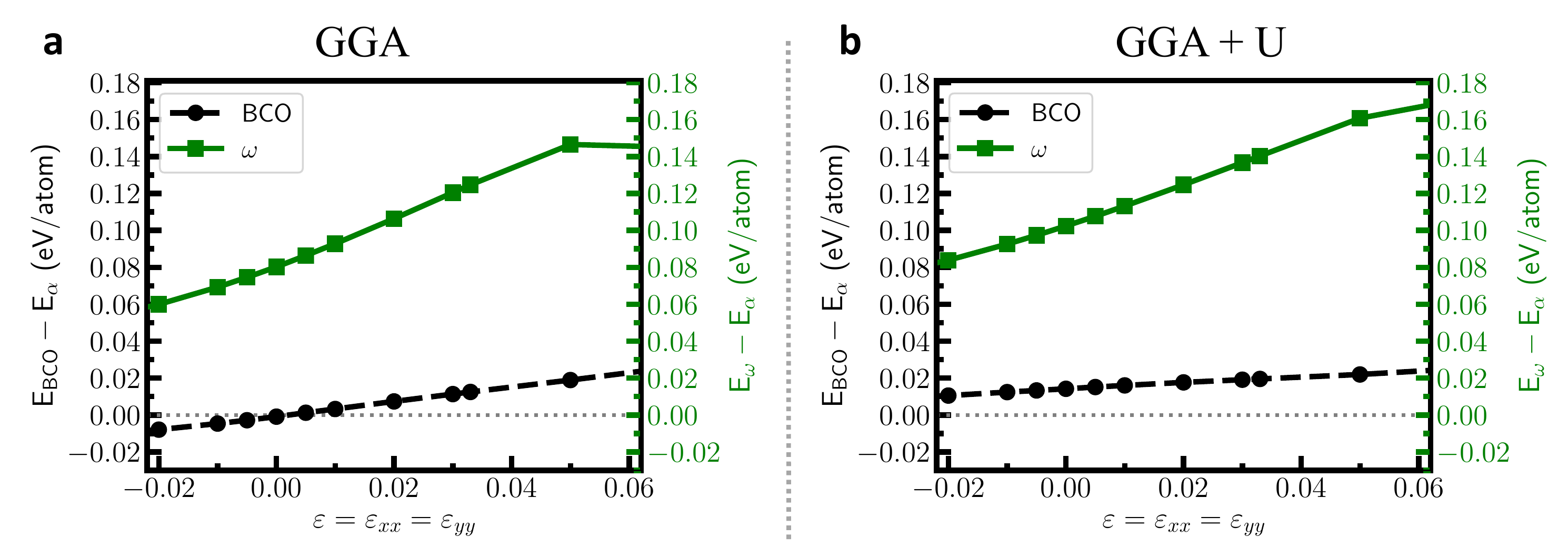}
\caption{The DFT calculated bulk energy differences (per atom) between the BCO and HCP (black circles) and $\omega$ and HCP (blue squares) phases of titanium as a function of applied biaxial strain ($\varepsilon = \varepsilon_{\text{xx}}=\varepsilon_{\text{yy}}$) using \textbf{a,} GGA (reproducing the results from \figsup{DFV}) and \textbf{b,} GGA+U exchange correlation functionals. The data point at zero applied strain corresponds to the strain state of the \tb twin boundary in which the BCO phase is under tensile stress imposed by the stress-free bulk HCP. }
  \label{figsup:GGAU_energies}
 \end{center}
 \end{figure}

 Using the GGA+U relaxed lattice constants, we then create the \tbBCO TB supercell (see Methods) and relax the structure.  \figsup{GGAU_TB} compares the relaxed structure of the \tbBCO TBs and shows that the morphology at the interface corresponds to the BCO structure, as found in the calculations with GGA (without Hubbard-U corrections). However, GGA+U predicts a narrower width compared to GGA. We understand this observation in the context of disjoining potential picture described above. Specifically, \figsup{GGAU_energies} compares the bulk free energy differences from GGA and GGA+U for the BCO and $\alpha$ as well as the $\omega$ and $\alpha$ Ti at the strain state imposed by the \tbBCO TB. The results show that GGA+U predicts increased (more positive) values for the BCO-HCP energy difference relative to GGA, with the HCP structure being stable at all strains. Within the disjoining potential picture explained above, this increase in bulk energy difference between BCO and HCP will favor a reduction in the volume of the higher-energy phase (BCO in this case), leading to a lower interfacial width, consistent with the supercell calculations.  
  
   \begin{table}[]
\caption{Relaxed lattice constants of $\alpha$, BCO, and $\omega$ phases of Ti using GGA and GGA+U exchange correlation functionals.  $\Delta E$ is the relative per atom energy of each phase relative to the $\alpha$ phase for each exchange correlation functional. Values in the parentheses correspond to the strained BCO under the biaxial strain state of the \tbBCO TB. }

\label{tabsup:DFTU}
\begin{adjustbox}{center}
\begin{tabular}{llccc}
         & Method & \multicolumn{1}{l}{$\text{a,b}(\AA)$}                                 & \multicolumn{1}{l}{$\text{c}(\AA)$} & \multicolumn{1}{l}{$\Delta E (\frac{\text{meV}}{\text{atom}})$} \\ \hline
$\alpha$ & GGA    & 2.936                                                                 & 4.647                               & 0                                                               \\
         & GGA+U  & 2.972                                                                 & 4.719                               & 0                                                               \\
BCO      & GGA    & \begin{tabular}[c]{@{}c@{}}7.448, 5.013\\ (7.489, 5.086)\end{tabular} & 5.585                               & \begin{tabular}[c]{@{}c@{}}-2.810\\ (-0.881)\end{tabular}       \\
         & GGA+U  & \begin{tabular}[c]{@{}c@{}}7.634, 5.066\\ (7.587, 5.146)\end{tabular} & 5.621                               & \begin{tabular}[c]{@{}c@{}}12.916\\ (14.202)\end{tabular}       \\
$\omega$ & GGA    & 4.576                                                                 & 2.829                               & -6.473                                                          \\
         & GGA+U  & 4.648                                                                 & 2.856                               & 7.197                                                           \\ \hline
\end{tabular}
\end{adjustbox}
\end{table}

\section{Effect of oxygen interstitials on the bulk phases and the \tbBCO twin boundary}
\label{secsup:oxygen}

DFT-GGA calculations were used to study the energetics of oxygen (O) impurity solutes in different bulk phases and in the vicinity of the \tbBCO TB in Ti. First, we compare the relative formation energy of oxygen atoms in $\alpha$, BCO, and $\omega$ phases. These calculations are also performed for the BCO structure at the biaxial strain state imposed by the \tbBCO TB. \tabsup{phases} lists the relaxed lattice parameters, space groups, and Wyckoff positions of lattice sites in these phases (for computational settings, see the Methods section). 

\begin{figure}[h!]
 \begin{center}
   \includegraphics[width=\textwidth]{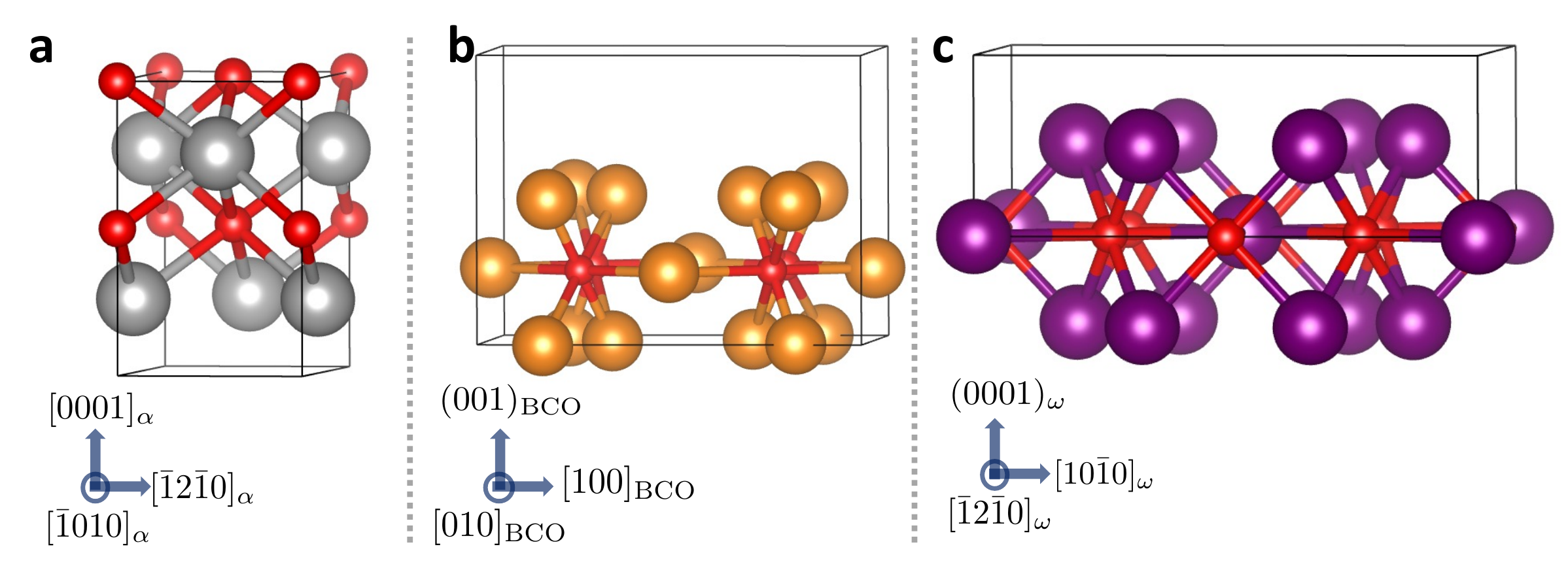}
\caption{Oxygen atoms at the lowest energy octahedral interstitial sites in: \textbf{a}, $\alpha$, \textbf{b}, BCO, and \textbf{c}, $\omega$ phases in Ti. Gray, orange and purple colors correspond to the Ti atoms in the $\alpha$, BCO and $\omega$ phases. Oxygen atoms are shown in red. }
  \label{figsup:oxygen_bulk}
 \end{center}
 \end{figure}

The stable interstitial sites in the BCO phase were found through a systematic search by placing an oxygen atom into the regions between Ti atoms and relaxing the atomic positions.  We found four distinct interstitial sites in the BCO phase (labelled as \textit{i}-\textit{iv}) among which the lowest energy site corresponds to the octahedral site occupying maximum interstitial volume between the host Ti atoms. Similarly, oxygen prefers to occupy an octahedral interstitial site surrounded by six Ti atoms in $\alpha$ and  $\omega$ phases. Other metastable sites include hexahedral and crowdion sites in $\alpha$ and hexahedral and tetrahedral sites in the $\omega$ phases  \cite{hennig2005}.  \figsup{oxygen_bulk} shows the location of stable octahedral interstitial sites for oxygen in $\alpha$, BCO, and $\omega$ phases. 

The formation energies for oxygen at different Wyckoff positions of the stable and metastable sites in the $\alpha$, BCO, and $\omega$ phases as well as the strained BCO complexion are calculated and listed in \tabsup{sites_bulk}. These energies are reported relative to molecular $\text{O}_2$. A comparison between the interstitial impurity formation energies among these phases shows that the octahedral formation energy is lower in $\alpha$. The difference between the site energies shifts the relative energy of crystal structures, decreasing the stability of BCO, and $\omega$ over $\alpha$. Additionally, this observation is consistent with the role of oxygen as an $\alpha$ stabilizer element reported in the literature \cite{trinkle2003}. The higher formation energy of oxygen in the strained state of BCO complexion found at the interface of the \tbBCO TB than the $\alpha$ phase indicates that oxygen atoms energetically prefer to stay at the bulk HCP sites compared to the strained BCO interfacial region. 

\begin{figure}[h!]
 \begin{center}
   \includegraphics[width=\textwidth]{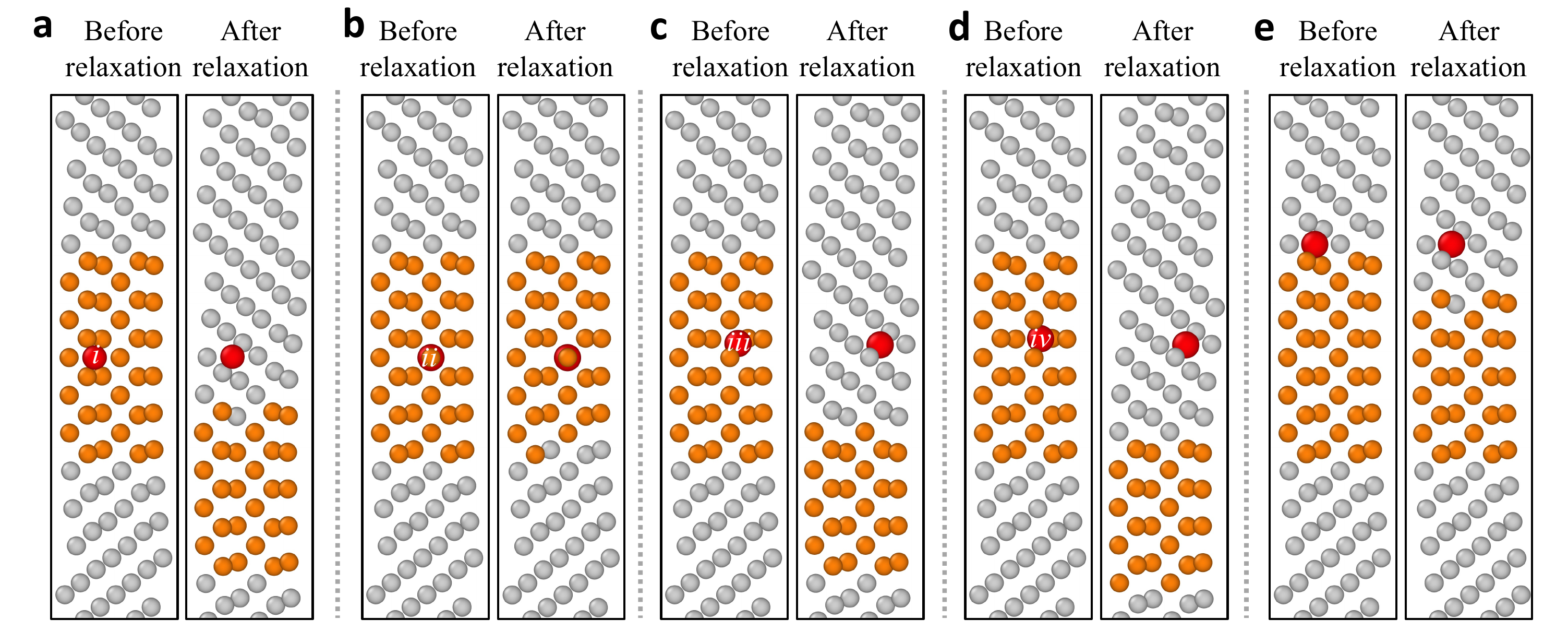}
\caption{Interaction of oxygen interstitials and the \tbBCO twin boundary in Ti. \textbf{a}-\textbf{d,} show the supercells with oxygen at different interstitial sties at the \tbBCO TB before and after relaxation. Labels \textit{i}-\textit{iv} correspond to the equivalent interstitial sites in the BCO bulk phase (see \tabsup{sites_bulk}).  \textbf{e,} shows the relaxation corresponding to site at the $\alpha$/BCO interface. Color codes are based on CNA parameter \cite{honeycutt1987} with gray representing HCP and orange showing defective regions. Oxygen atoms are shown in red. }
  \label{figsup:oxygen_twin}
 \end{center}
 \end{figure}

Next, we explore the effect of oxygen on the \tbBCO TB by directly computing the interaction energy between oxygen and the TB using DFT. Oxygen/TB interaction energies can be defined as the difference between the energy of the supercell with oxygen located at the corresponding interstitial site and that of a supercell with oxygen at the reference octahedral site in the bulk-like HCP region. \figsup{oxygen_twin}a-d show the supercells with oxygen atoms placed at different interstitial sites at the interface of \tbBCO TB before and after energy minimization. \figsup{oxygen_twin}e shows the case where the oxygen atom is placed at the interface of BCO complexion and the HCP phase. The results demonstrate that it is energetically unfavorable for oxygen atoms to stay at the interfacial sites. Due to the large repulsive interaction between oxygen atoms and the TB, the interface is pushed away by few layers during the energy relaxation calculation, to enable the oxygen atom to be located in a bulk HCP octahedral environment as highlighted in \figsup{oxygen_twin}a, c, and d (corresponding to the sites \textit{i}, \textit{iii}, and \textit{iv} in the bulk phase). While the interface remains stable in the case of \figsup{oxygen_twin}b  interstitial site (corresponding to the site \textit{ii} in the bulk phase), there is a large repulsive interaction energy of $0.921 \, \text{eV}$ suggesting that it is energetically favorable for oxygen atoms to stay in the bulk over the \tbBCO TB. 

While the oxygen atoms are disfavored from segregating to the \tbBCO TB, they act as a hardening agent leading to the increase in the yield stress \cite{chong2020}. Therefore, the hardening induced by oxygen interstitials can create high enough stress levels to activate the \tbBCO twins which are not observed in the lower oxygen content samples. We believe this effect is the dominant one in correlating the presence of oxygen with the \tb TB formation, rather than being associated with a segregation-induced interface structure transition, due to the highly repulsive interactions between oxygen and the TB.  

The results above also provide insights into why the DFT calculations for of the TB width in pure Ti are larger than those observed experimentally in the samples with oxygen.  Specifically, as discussed in the main text and above, if the disjoining potential varies weakly with oxygen content, the role of oxygen in increasing the energy of BCO relative to HCP should lead to a narrowing of the interface width.  To explore this interpretation further, our findings here show that the oxygen concentration of $0.3$ wt.\% ($0.88$ at.\%) raises the energy difference between (strained) BCO and HCP structures by roughly $8 \, \text{meV}$ which is the same energy increase realized when pure Ti is biaxially strained by $1.5\, \%$ applied tensile strain (see \figsup{DFV}). This amount of energy penalty leads to a narrower interfacial width of $11.07 \, \AA $ (see \fig{transitions}b), which is in a good agreement with the $11.70 \, \AA $ width measured experimentally.

   \begin{table}[]
\caption{Relaxed structural parameters and space groups for $\alpha$, BCO, and $\omega$-Ti. Wyckoff positions of lattice sites within each phase are also listed. For the relaxed BCO structure, the ideal position of 8(j) Wyckoff lattice site is at $x=0.414$ and $y=0.677$. Subscripts A and B refers to the two atoms per unit cell of BCO and $\omega$ phases.    }

\label{tabsup:phases}
\begin{adjustbox}{center}
\begin{tabular}{ll}
Site                                & Wyckoff position                                                    \\ \hline
\multicolumn{2}{l}{a) $\alpha$ phase ($a=2.94 \, \AA$, $c=4.65 \, \AA$); space group: $P6_3/mmc$}         \\
$\alpha$                            & $2\text{(c)} \, (\frac{1}{3},\frac{2}{3},\frac{1}{4})$              \\ \hline
\multicolumn{2}{l}{b) BCO phase ($a=7.44 \, \AA$, $b=5.02 \, \AA$, $c=5.58 \, \AA$); space group: $Ibam$} \\
$\text{BCO}_\text{A}$               & $4\text{(b)} \, (\frac{1}{2},0,\frac{1}{4})$                        \\
$\text{BCO}_\text{B}$               & $8\text{(j)} \, (x,y,0)$                                            \\ \hline
\multicolumn{2}{l}{c) $\omega$ phase ($a=4.58 \, \AA$, $c=2.83 \, \AA$); space group: $P6/mmm$}           \\
$\omega_\text{A}$                   & $1\text{(a)} \, (0,0,0)$                                            \\
$\omega_\text{B}$                   & $2\text{(d)} \, (\frac{1}{3},\frac{2}{3},\frac{1}{2})$              \\ \hline
\end{tabular}
\end{adjustbox}
\end{table}

   \begin{table}[]
\caption{Formation energies and Wyckoff position of the minimum energy oxygen interstitial in the $\alpha$, BCO, strained BCO and $\omega$ phase derived from DFT-GGA calculations. The subscripts oct, hex, crowd and tet refer to the octahedral, hexahedral, crowdion, and tetrahedral interstitial positions. The formation energies $\text{E}_\text{f}$ are measured relative to molecular $\text{O}_2$. The interstitial site energies $\Delta E$ are reported relative to the octahedral site energy in the $\alpha$ phase. Values in parentheses correspond to the site energies at the BCO under the biaxial strain state of the \tbBCO TB . Calculations are performed at $1 \, \%$ oxygen concentration. The coordination number $Z$ for each site is also included. In the BCO phase, ideal interstitial positions are at $x=0.796$, $y=0.086$, $z=0.125$ for the 16(k) and $x=0.225$, $y=0.565$ for the 8(j) sites, respectively. In the $\omega$ phase, ideal $x$ is $0.240$, and $0.120$ for the 6(k) and 6(m) lattice sites, respectively.  }

\label{tabsup:sites_bulk}
\begin{adjustbox}{center}

\begin{tabular}{lllcc}
Site                                 & Wyckoff position                                       & Z & $\text{E}_\text{f}$ (eV) & $\Delta E$ (eV)    \\ \hline
\multicolumn{5}{l}{(a) $\alpha$ interstitial sites}                                                                                               \\
$\alpha_\text{oct}$                  & $2\text{(a)} \, (0,0,0)$                               & 6 & $-6.24$                  & $+0.00$            \\
$\alpha_\text{hex}$                  & $2\text{(d)} \, (\frac{2}{3},\frac{1}{3},\frac{1}{4})$ & 5 & $-5.12$                  & $+1.12$            \\
$\alpha_\text{crowd}$                & $6\text{(g)} \, (\frac{1}{2},0,0)$                     & 6 & $-4.52$                  & $+1.72$            \\ \hline
\multicolumn{5}{l}{(b) BCO interstitial sites}                                                                                                    \\
$\text{BCO}_\text{\textit{i}}$ (oct) & $8\text{(e)} \, (\frac{1}{4},\frac{3}{4},\frac{1}{4})$ & 6 & $-5.71 \, (-5.67)$       & $+0.53 \, (+0.57)$ \\
$\text{BCO}_\text{\textit{ii}}$      & $4\text{(a)} \, (\frac{1}{2},\frac{3}{4},\frac{3}{4})$ & 4 & $-5.41 \, (-5.37)$       & $+0.83 \, (+0.87)$ \\
$\text{BCO}_\text{\textit{iii}}$     & $16\text{(k)} \, (x,y,z)$                              & 5 & $-5.25 \, (-5.32)$       & $+0.99 \, (+0.92)$ \\
$\text{BCO}_\text{\textit{iv}}$      & $8\text{(j)} \, (x,y,0)$                               & 5 & $-5.08 \, (-5.16)$       & $+1.16 \, (+1.08)$ \\ \hline
\multicolumn{5}{l}{(c) $\omega$ interstitial sites}                                                                                               \\
$\omega_\text{oct}$                  & $3\text{(f)} \, (\frac{1}{2},0,0)$                     & 6 & $-6.19$                  & $+0.05$           \\
$\omega_\text{hex}$                  & $6\text{(m)} \, (x,2x,\frac{1}{2})$                    & 5 & $-4.59$                  & $+1.65$            \\
$\omega_\text{tet}$                  & $6\text{(k)} \, (x,0,\frac{1}{2})$                     & 4 & $-4.45$                  & $+1.79$            \\ \hline
\end{tabular}

\end{adjustbox}
\end{table}

\pagebreak
\newpage
\clearpage



 \section*{References}
\bibliographystyle{unsrt}

%% file: main.bbl
\begin{thebibliography}{10}

\bibitem{christian1995}
J.~W. Christian and S.~Mahajan.
\newblock Deformation twinning.
\newblock {\em Progress in Materials Science}, 39(1-2):1--157, 1995.

\bibitem{sutton1995}
A.~P. Sutton.
\newblock Interfaces in crystalline materials.
\newblock {\em Monographs on the Physice and Chemistry of Materials}, pages
  414--423, 1995.

\bibitem{kaplan2013}
W.~D. Kaplan, D.~Chatain, P.~Wynblatt, and W.~C. Carter.
\newblock A review of wetting versus adsorption, complexions, and related
  phenomena: The rosetta stone of wetting.
\newblock {\em Journal of Materials Science}, 48(17):5681--5717, September
  2013.

\bibitem{mishin2010}
Y.~Mishin, M.~Asta, and J.~Li.
\newblock Atomistic modeling of interfaces and their impact on microstructure
  and properties.
\newblock {\em Acta Materialia}, 58(4):1117--1151, February 2010.

\bibitem{frolov2015}
T.~Frolov and Y.~Mishin.
\newblock Phases, phase equilibria, and phase rules in low-dimensional systems.
\newblock {\em The Journal of Chemical Physics}, 143(4):044706, July 2015.

\bibitem{frolov2012}
T.~Frolov and Y.~Mishin.
\newblock Thermodynamics of coherent interfaces under mechanical stresses.
  {{I}}. {{Theory}}.
\newblock {\em Physical Review B}, 85(22):224106, June 2012.

\bibitem{hart1972}
E.~W. Hart.
\newblock Grain boundary phase transformations.
\newblock In {\em The Nature and Behavior of Grain Boundaries}, pages 155--170.
  {Springer US}, {New York, NY}, 1972.

\bibitem{cahn1982}
J.~W. Cahn.
\newblock Transitions and phase equilibria among grain boundary structures.
\newblock {\em Le Journal de Physique Colloques}, 43(C6):C6--199--C6--213,
  December 1982.

\bibitem{cantwell2020}
P.~R. Cantwell, T.~Frolov, T.~J. Rupert, A.~R. Krause, C.~J. Marvel, G.~S.
  Rohrer, J.~M. Rickman, and M.~P. Harmer.
\newblock Grain boundary complexion transitions.
\newblock {\em Annual Review of Materials Research}, 50(1):465--492, July 2020.

\bibitem{meiners2020}
T.~Meiners, T.~Frolov, R.~E. Rudd, G.~Dehm, and C.~H. Liebscher.
\newblock Observations of grain-boundary phase transformations in an elemental
  metal.
\newblock {\em Nature}, 579(7799):375--378, March 2020.

\bibitem{frolovStructuralPhaseTransformations2013}
T.~Frolov, D.~L. Olmsted, M.~Asta, and Y.~Mishin.
\newblock Structural phase transformations in metallic grain boundaries.
\newblock {\em Nature Communications}, 4(1):1899, October 2013.

\bibitem{frolov2013}
T.~Frolov, S.~V. Divinski, M.~Asta, and Y.~Mishin.
\newblock Effect of interface phase transformations on diffusion and
  segregation in high-angle grain boundaries.
\newblock {\em Physical Review Letters}, 110(25):255502, June 2013.

\bibitem{tang2006a}
M.~Tang, W.~C. Carter, and R.~M. Cannon.
\newblock Diffuse interface model for structural transitions of grain
  boundaries.
\newblock {\em Physical Review B}, 73(2):024102, January 2006.

\bibitem{dillon2007}
S.~J. Dillon, M.~Tang, W.~C. Carter, and M.~P. Harmer.
\newblock Complexion: {{A}} new concept for kinetic engineering in materials
  science.
\newblock {\em Acta Materialia}, 55(18):6208--6218, October 2007.

\bibitem{hart1968}
E.~W. Hart.
\newblock Two-dimensional phase transformation in grain boundaries.
\newblock {\em Scripta Metallurgica}, 2(3):179--182, March 1968.

\bibitem{luo2016}
J.~Luo.
\newblock A short review of high-temperature wetting and complexion transitions
  with a critical assessment of their influence on liquid metal embrittlement
  and corrosion.
\newblock {\em Corrosion}, 72(7):897--910, July 2016.

\bibitem{cantwell2014}
P.~R. Cantwell, M.~Tang, S.~J. Dillon, J.~Luo, G.~S. Rohrer, and M.~P. Harmer.
\newblock Grain boundary complexions.
\newblock {\em Acta Materialia}, 62:1--48, January 2014.

\bibitem{gibbs1948}
J.~W. Gibbs.
\newblock {\em The Collected Works of {{J}}. {{Willard Gibbs}}}.
\newblock {Yale University Press}, 1948.

\bibitem{frolov2014}
T.~Frolov.
\newblock Effect of interfacial structural phase transitions on the coupled
  motion of grain boundaries: {{A}} molecular dynamics study.
\newblock {\em Applied Physics Letters}, 104(21):211905, May 2014.

\bibitem{duscher2004}
G.~Duscher, M.~F. Chisholm, U.~Alber, and M.~R{\"u}hle.
\newblock Bismuth-induced embrittlement of copper grain boundaries.
\newblock {\em Nature Materials}, 3(9):621--626, September 2004.

\bibitem{frolov2015a}
T.~Frolov, M.~Asta, and Y.~Mishin.
\newblock Segregation-induced phase transformations in grain boundaries.
\newblock {\em Physical Review B}, 92(2):020103, July 2015.

\bibitem{frolov2016}
T.~Frolov, M.~Asta, and Y.~Mishin.
\newblock Phase transformations at interfaces: {{Observations}} from atomistic
  modeling.
\newblock {\em Current Opinion in Solid State and Materials Science},
  20(5):308--315, October 2016.

\bibitem{frolov2018}
T.~Frolov, W.~Setyawan, R.~J. Kurtz, J.~Marian, A.~R. Oganov, R.~E. Rudd, and
  Q.~Zhu.
\newblock Grain boundary phases in bcc metals.
\newblock {\em Nanoscale}, 10(17):8253--8268, 2018.

\bibitem{olmsted2011}
D.~L. Olmsted, D.~Buta, A.~Adland, S.~M. Foiles, M.~Asta, and A.~Karma.
\newblock Dislocation-pairing transitions in hot grain boundaries.
\newblock {\em Physical Review Letters}, 106(4):046101, January 2011.

\bibitem{mendelev2013}
M.~Mendelev and A.~King.
\newblock The interactions of self-interstitials with twin boundaries.
\newblock {\em Philosophical Magazine}, 93(10-12):1268--1278, April 2013.

\bibitem{chong2020}
Y.~Chong, M.~Poschmann, R.~Zhang, S.~Zhao, M.~S. Hooshmand, E.~Rothchild, D.~L.
  Olmsted, J.~W. Morris, D.~C. Chrzan, M.~Asta, and A.~M. Minor.
\newblock Mechanistic basis of oxygen sensitivity in titanium.
\newblock {\em Science Advances}, 6(43):eabc4060, October 2020.

\bibitem{wangAtomicStructuresSymmetric2012}
J.~Wang and I.~J. Beyerlein.
\newblock Atomic structures of symmetric tilt grain boundaries in hexagonal
  close packed (hcp) crystals.
\newblock {\em Modelling and Simulation in Materials Science and Engineering},
  20(2):024002, March 2012.

\bibitem{honeycutt1987}
J.~D. Honeycutt and H.~C. Andersen.
\newblock Molecular dynamics study of melting and freezing of small
  {{Lennard}}-{{Jones}} clusters.
\newblock {\em The Journal of Physical Chemistry}, 91(19):4950--4963, September
  1987.

\bibitem{zarkevich2016}
N.~A. Zarkevich and D.~D. Johnson.
\newblock Titanium {$\alpha$} - {$\omega$} phase transformation pathway and a
  predicted metastable structure.
\newblock {\em Physical Review B}, 93(2):020104, January 2016.

\bibitem{zhu2018}
Q.~Zhu, A.~Samanta, B.~Li, R.~E. Rudd, and T.~Frolov.
\newblock Predicting phase behavior of grain boundaries with evolutionary
  search and machine learning.
\newblock {\em Nature Communications}, 9(1):467, December 2018.

\bibitem{oganov2006}
A.~R. Oganov and C.~W. Glass.
\newblock Crystal structure prediction using ab initio evolutionary techniques:
  {{Principles}} and applications.
\newblock {\em The Journal of Chemical Physics}, 124(24):244704, June 2006.

\bibitem{lyakhov2013}
A.~O. Lyakhov, A.~R. Oganov, H.~T. Stokes, and Q.~Zhu.
\newblock New developments in evolutionary structure prediction algorithm
  {{USPEX}}.
\newblock {\em Computer Physics Communications}, 184(4):1172--1182, April 2013.

\bibitem{hennig2008}
R.~G. Hennig, T.~J. Lenosky, D.~R. Trinkle, S.~P. Rudin, and J.~W. Wilkins.
\newblock Classical potential describes martensitic phase transformations
  between the {$\alpha$} , {$\beta$} , and {$\omega$} titanium phases.
\newblock {\em Physical Review B}, 78(5):054121, August 2008.

\bibitem{lipowsky1987}
R.~Lipowsky and M.~E. Fisher.
\newblock Scaling regimes and functional renormalization for wetting
  transitions.
\newblock {\em Physical Review B}, 36(4):2126--2141, August 1987.

\bibitem{fisher1985}
D.~S. Fisher and D.~A. Huse.
\newblock Wetting transitions: {{A}} functional renormalization-group approach.
\newblock {\em Physical Review B}, 32(1):247--256, July 1985.

\bibitem{Zhang2017}
J.~Zhang, C.~C. Tasan, M.~J. Lai, A.~C. Dippel, and D.~Raabe.
\newblock Complexion-mediated martensitic phase transformation in {{Titanium}}.
\newblock {\em Nature Communications}, 8(1):14210, April 2017.

\bibitem{medlin2001}
D.~Medlin, S.~Foiles, and D.~Cohen.
\newblock A dislocation-based description of grain boundary dissociation:
  Application to a 90\textdegree{} {$\langle$}110{$\rangle$} tilt boundary in
  gold.
\newblock {\em Acta Materialia}, 49(18):3689--3697, October 2001.

\bibitem{lukas2007}
H.~Lukas, S.~G. Fries, and B.~Sundman.
\newblock {\em Computational Thermodynamics: The {{Calphad}} Method}.
\newblock {Cambridge university press}, 2007.

\bibitem{kresse1996}
G.~Kresse and J.~Furthm{\"u}ller.
\newblock Efficient iterative schemes for ab initio total-energy calculations
  using a plane-wave basis set.
\newblock {\em Physical Review B}, 54(16):11169--11186, October 1996.

\bibitem{kresse1999}
G.~Kresse and D.~Joubert.
\newblock From ultrasoft pseudopotentials to the projector augmented-wave
  method.
\newblock {\em Physical review b}, 59(3):1758, 1999.

\bibitem{perdew1992}
J.~P. Perdew and Y.~Wang.
\newblock Accurate and simple analytic representation of the electron-gas
  correlation energy.
\newblock {\em Physical review B}, 45(23):13244, 1992.

\bibitem{methfessel1989}
M.~Methfessel and A.~T. Paxton.
\newblock High-precision sampling for {{Brillouin}}-zone integration in metals.
\newblock {\em Physical Review B}, 40(6):3616, 1989.

\bibitem{plimpton1995}
S.~Plimpton.
\newblock Fast parallel algorithms for short-range molecular dynamics.
\newblock {\em Journal of Computational Physics}, 117(1):1--19, March 1995.

\bibitem{sheppard2012}
D.~Sheppard, P.~Xiao, W.~Chemelewski, D.~D. Johnson, and G.~Henkelman.
\newblock A generalized solid-state nudged elastic band method.
\newblock {\em The Journal of Chemical Physics}, 136(7):074103, February 2012.

\bibitem{henkelman2015}
G.~Henkelman.
\newblock {{VTST}} tools.
\newblock {\em Available from: http://theory.cm.utexas.edu/vtsttools/.}, 2015.

\bibitem{savitzky2021}
B.~H. Savitzky, S.~E. Zeltmann, L.~A. Hughes, H.~G. Brown, S.~Zhao, P.~M. Pelz,
  T.~C. Pekin, E.~S. Barnard, J.~Donohue, L.~Rangel~DaCosta, E.~Kennedy,
  Y.~Xie, M.~T. Janish, M.~M. Schneider, P.~Herring, C.~Gopal, A.~Anapolsky,
  R.~Dhall, K.~C. Bustillo, P.~Ercius, M.~C. Scott, J.~Ciston, A.~M. Minor, and
  C.~Ophus.
\newblock {{py4DSTEM}}: {{A Software Package}} for {{Four}}-{{Dimensional
  Scanning Transmission Electron Microscopy Data Analysis}}.
\newblock {\em Microscopy and Microanalysis}, pages 1--32, May 2021.

\bibitem{ophus2017}
C.~Ophus.
\newblock A fast image simulation algorithm for scanning transmission electron
  microscopy.
\newblock {\em Advanced Structural and Chemical Imaging}, 3(1):13, December
  2017.

\bibitem{pryor2017}
A.~Pryor, C.~Ophus, and J.~Miao.
\newblock A streaming multi-{{GPU}} implementation of image simulation
  algorithms for scanning transmission electron microscopy.
\newblock {\em Advanced Structural and Chemical Imaging}, 3(1):15, December
  2017.

\end{thebibliography}

\begin{thebibliography}{10}

\bibitem{lipowsky1987}
Reinhard Lipowsky and Michael~E. Fisher.
\newblock Scaling regimes and functional renormalization for wetting
  transitions.
\newblock {\em Physical Review B}, 36(4):2126--2141, August 1987.

\bibitem{fisher1985}
Daniel~S. Fisher and David~A. Huse.
\newblock Wetting transitions: {{A}} functional renormalization-group approach.
\newblock {\em Physical Review B}, 32(1):247--256, July 1985.

\bibitem{minonishi1982}
Y.~Minonishi, S.~Ishioka, M.~Koiwa, and S.~Mobozumi.
\newblock The structure of \{11-21\} twin boundaries in {{HCP}} crystals.
\newblock {\em physica status solidi (a)}, 71(1):253--258, May 1982.

\bibitem{dudarev1998}
S.~L. Dudarev, G.~A. Botton, S.~Y. Savrasov, C.~J. Humphreys, and A.~P. Sutton.
\newblock Electron-energy-loss spectra and the structural stability of nickel
  oxide: {{An LSDA}}+{{U}} study.
\newblock {\em Physical Review B}, 57(3):1505--1509, January 1998.

\bibitem{lutfalla2011}
Suzanne Lutfalla, Vladimir Shapovalov, and Alexis~T. Bell.
\newblock Calibration of the {{DFT}}/{{GGA}}+{{U Method}} for {{Determination}}
  of {{Reduction Energies}} for {{Transition}} and {{Rare Earth Metal Oxides}}
  of {{Ti}}, {{V}}, {{Mo}}, and {{Ce}}.
\newblock {\em Journal of Chemical Theory and Computation}, 7(7):2218--2223,
  July 2011.

\bibitem{zarkevich2016}
N.~A. Zarkevich and D.~D. Johnson.
\newblock Titanium {$\alpha$} - {$\omega$} phase transformation pathway and a
  predicted metastable structure.
\newblock {\em Physical Review B}, 93(2):020104, January 2016.

\bibitem{honeycutt1987}
J.~Dana. Honeycutt and Hans~C. Andersen.
\newblock Molecular dynamics study of melting and freezing of small
  {{Lennard}}-{{Jones}} clusters.
\newblock {\em The Journal of Physical Chemistry}, 91(19):4950--4963, September
  1987.

\bibitem{hennig2005}
Richard~G. Hennig, Dallas~R. Trinkle, Johann Bouchet, Srivilliputhur~G.
  Srinivasan, Robert~C. Albers, and John~W. Wilkins.
\newblock Impurities block the {$\alpha$} to {$\omega$} martensitic
  transformation in titanium.
\newblock {\em Nature Materials}, 4(2):129--133, February 2005.

\bibitem{trinkle2003}
D.~R. Trinkle, R.~G. Hennig, S.~G. Srinivasan, D.~M. Hatch, M.~D. Jones, H.~T.
  Stokes, R.~C. Albers, and J.~W. Wilkins.
\newblock New mechanism for the {$\alpha$} to {$\omega$} martensitic
  transformation in pure titanium.
\newblock {\em Physical Review Letters}, 91(2):025701, July 2003.

\bibitem{chong2020}
Yan Chong, Max Poschmann, Ruopeng Zhang, Shiteng Zhao, Mohammad~S. Hooshmand,
  Eric Rothchild, David~L. Olmsted, J.~W. Morris, Daryl~C. Chrzan, Mark Asta,
  and Andrew~M. Minor.
\newblock Mechanistic basis of oxygen sensitivity in titanium.
\newblock {\em Science Advances}, 6(43):eabc4060, October 2020.

\end{thebibliography}
